\documentclass[iop]{emulateapj}
\usepackage{graphics}
\usepackage{graphicx}
\usepackage{epstopdf}
\usepackage{epsfig}
\usepackage{hyperref}

\begin{document}

\title{Modeling Multi-Wavelength Stellar Astrometry. I. SIM Lite Observations of Interacting Binaries}

\author{Jeffrey L. Coughlin\altaffilmark{1,6}, Dawn M. Gelino\altaffilmark{2}, Thomas E. Harrison\altaffilmark{1}, D. W. Hoard\altaffilmark{3}, David R. Ciardi\altaffilmark{2}, G. Fritz Benedict\altaffilmark{4}, Steve B. Howell\altaffilmark{5}, Barbara E. McArthur\altaffilmark{4}, \& Stefanie Wachter\altaffilmark{3}}

\altaffiltext{1}{Department of Astronomy, New Mexico State University, P.O. Box 30001, MSC 4500, Las Cruces, New Mexico 88003-8001; jlcough@nmsu.edu}
\altaffiltext{2}{NASA Exoplanet Science Institute, California Institute of Technology, Pasadena, CA 91125}
\altaffiltext{3}{Spitzer Science Center, California Institute of Technology, Pasadena, CA 91125}
\altaffiltext{4}{McDonald Observatory, University of Texas, Austin, Texas 78712}
\altaffiltext{5}{NOAO, 950 N. Cherry Avenue, Tucson, AZ 85719}
\altaffiltext{6}{NSF Graduate Research Fellow}

\slugcomment{Accepted to the Astrophysical Journal}

\begin{abstract}
Interacting binaries consist of a secondary star which fills or is very close to filling its Roche lobe, resulting in accretion onto the primary star, which is often, but not always, a compact object. In many cases, the primary star, secondary star, and the accretion disk can all be significant sources of luminosity. SIM Lite will only measure the photocenter of an astrometric target, and thus determining the true astrometric orbits of such systems will be difficult. We have modified the Eclipsing Light Curve code \citep{Orosz00} to allow us to model the flux-weighted reflex motions of interacting binaries, in a code we call \textsc{reflux}. This code gives us sufficient flexibility to investigate nearly every configuration of interacting binary. We find that SIM Lite will be able to determine astrometric orbits for all sufficiently bright interacting binaries where the primary or secondary star dominates the luminosity. For systems where there are multiple components that comprise the spectrum in the optical bandpass accessible to SIM Lite, we find it is possible to obtain absolute masses for both components, although multi-wavelength photometry will be required to disentangle the multiple components. In all cases, SIM Lite will at least yield accurate inclinations, and provide valuable information that will allow us to begin to understand the complex evolution of mass-transferring binaries. It is critical that SIM Lite maintains a multi-wavelength capability to allow for the proper deconvolution of the astrometric orbits in multi-component systems.
\end{abstract}


\section{Introduction}

The field of stellar evolution has matured to the point that we generally understand the entire evolution of an isolated star if we know its initial mass. While many details remain to be solved, it is unlikely that SIM Lite will provide a revolution in our understanding of the evolutionary processes for single stars of low and intermediate mass. The same cannot be said for binary star systems, where there are at present a number of outstanding issues that remain unsolved. Among the most difficult is how mass transfer proceeds during the common envelope stage that occurs in the post-main sequence lives of close binaries. The formation of nearly every mass transferring binary requires a common envelope phase where the secondary star helps to strip the atmosphere of the primary while driving mass from the system, and shrinking the binary's orbit. Current theory \citep[c.f.,][]{Pod03} has a difficult time explaining the creation of systems with massive black hole primaries (M$_{\rm BH}$ $\geq$ 9 M$_{\sun}$) and very low mass secondary stars (M$_{\rm 2}$ $\leq$ 0.7 M$_{\sun}$). Similar problems exist across the mass spectrum of close and interacting binaries (IBs), which for the purposes of the current study include systems with a black hole, neutron star, or white dwarf (WD) primary and a non-degenerate companion. For example, \citet{Howell01b} explores the well known 2-3 hr period gap observed in cataclysmic variable (CV) systems, which is postulated to be a result of cessation of mass-transfer from the main-sequence, low-mass secondary star to the white dwarf companion for periods between 2-3 hours. \citet{Howell01b} runs binary population synthesis models that show this theory is only correct if the secondary stars in these systems are up to 50\% oversized, and thus both components are really much less massive than has been typically assumed via standard main-sequence mass-radius-temperature relations. Thus, if accurate masses can be determined for these systems, their formation theories can be directly tested.

Typically in binary star work, radial velocity curves are used to obtain masses for the components. While it can be quite simple to get a radial velocity curve for one component of an IB, (the secondary star if it dominates the luminosity, else typically the disk), it is difficult, or even impossible, to get such data for their black hole, neutron star, or WD primaries. Thus, along with the radial velocity amplitude of the secondary or disk, one must also assume a mass and know the binary orbital inclination in order to solve for the component masses. It is especially difficult to determine the binary inclination. In a large number of IBs the secondary star can be quite prominent at infrared wavelengths, and since the object is distorted, it exhibits ellipsoidal variations as it orbits the primary \citep[c.f.][]{Gelino01}. The amplitude of these variations are dependent on the orbital inclination, and to a lesser extent, the properties of the secondary star. However, even a small amount of contamination by the accretion disk introduces considerable uncertainty in this method, and nearly all interacting binary systems have some level of contamination from the accretion process. Thus, this technique of using infrared ellipsoidal variations to determine the inclination fails if one cannot ascertain the spectrum and level of the contaminating source.

This paper will show how SIM Lite, with multi-wavelength, microarcsecond astrometry, will be the first mission capable of determining the multi-wavelength astrometric orbits of IBs. The field of optical astrometry has recently been making great progress from both ground-based observations, as well as is posed to make dramatic leaps with the upcoming space-based missions GAIA and SIM Lite. The limits of ground-based astrometry have been recently pushed by both CHARA and PRIMA/VLTI, but neither are capable of the kind of astrometric measurements needed to probe IBs. CHARA has multi-wavelength capabilities, but can only provide angular resolution to $\sim$200 $\mu$as \citep{Brummelaar05}, which is far larger than the reflex motions to be discussed in this paper, and requires very bright targets. PRIMA/VLTI will achieve $\sim$30-40 $\mu$as precision \citep{Belle2008}, which may be at the limit of usefulness for these systems, but only in K-band, which does not help distinguish individual component orbits, as will be shown. The GAIA mission will provide astrometry for $\sim$10$^{9}$ objects with 4 - 160 $\mu$as accuracy, for stars with V=10-20 respectively, and does posses multi-wavelength capabilities \citep{Cacciari09}. However, GAIA is a scanning satellite and cannot perform pointed or time-critical observations, and in fact only achieves this accuracy by averaging $\sim$80 individual measurements, the individual errors of which range from 36 - 1,431 $\mu$as, again for stars with V=10-20 respectively \citep{Mignard05}. Thus, GAIA cannot provide the high-precision, time-critical pointed observations needed to study IBs. In contrast, SIM Lite will be able to point at any desired object for any length of time, providing single-measurement accuracy of $\sim$1 $\mu$as \citep{NASA09}. As well, SIM Lite will have $\sim$80 spectral channels, spanning 450 to 900 nm, thus providing multi-wavelength, microarcsecond astrometry \citep{NASA09}.

SIM Lite will offer the first chance at measuring astrometric orbits for a large number of IBs, thus directly yielding inclinations and allowing for the precise measurement of the masses of both components in these systems. Determining the inclination of a zero-eccentricity system from astrometry is as straightforward as determining the ratio the semi-major and semi-minor axes of the projected ellipse on the sky. Ideally, one can derive accurate values for the masses of both components from the astrometry of a single component \citep{Benedict00}, but to do so requires that one knows the orbital period, the semi-major axis of the apparent orbit, the parallax of the system, and the mass ratio of the two components. SIM Lite can provide the first three of these four parameters, but the mass ratios must be estimated for many of the main systems of interest. However, for binaries where there is a significant amount of light from more than one component, \emph{SIM Lite can determine absolute masses for both components}. Since SIM Lite only measures the location of the photocenter of the system, the observed reflex motion of the system will be wavelength-dependent. For example, in a binary where one component is hotter than the other, the measured motion in the blue part of the spectrum will be different from that measured in the red, with shorter wavelengths tracing the motion of the hotter component, and vice-versa. Thus, if one has at least two astrometric orbits at different wavelengths, and one knows the ratio of luminosities in each bandpass, (i.e. the spectra of the components), then one could reconstruct the individual orbits of each component and obtain absolute masses for both. Since SIM Lite is currently designed to have $\sim$80 spectral channels, and since multi-color photometry and spectroscopy already exists for most IBs of interest, this technique should be quite feasible for most IBs.

In the next section we describe the procedure and code used to model the reflex motions of IBs. In section 3 we examine the results for ``proto-typical'' systems, including both X-ray binaries and cataclysmic variables, and estimate the required observing time required by SIM Lite to obtain reasonably accurate parameters for each system. We also briefly investigate how the presence of disk temperature gradients and hotspots in the systems affect the derived orbits. In section 4 we present a full modeling of simulated SIM Lite data for one system, and show the precision and accuracy of recovered astrometric parameters and derived system masses. We summarize our results in section 5.

\section{The Modeling Procedure: The {\sc reflux} Code}

\textsc{reflux}\footnotemark[1] is a code that computes the flux-weighted astrometric reflex motions of binary systems. At its core is the Eclipsing Light Curve (ELC) code, which is normally used to compute light curves of eclipsing binary systems (Orosz \& Hauschildt 2000). Besides specifying the primary and secondary stars, ELC allows for the inclusion of an accretion disk.  As with other light curve modeling programs, an array of physical effects are taken into account, such as non-spherical geometry, gravity brightening, limb darkening, mutual heating, and reflection effects. The program can either use a blackbody formula for local intensities of the stellar components, or interpolate from a large grid of NextGen model atmospheres (Hauschildt et al. 1999). ELC also allows for up to two hot or cool spots to be placed on each star, and on the accretion disk. Thus, ELC can reproduce nearly any binary system, including complicated systems such as cataclysmic variables, RS CVn systems, and Low Mass X-ray Binaries (LMXBs).  Additionally, we have modified the ELC code to allow for a mixture of blackbody and model atmosphere intensities for use among different system components, and to allow for the possibility of a free-free, or bremsstrahlung, accretion disk, which follows the form $F_{\lambda} \propto T^{-\frac{1}{2}}e^{-\frac{hc}{\lambda kT}}\lambda^{-2}$, where T is the temperature in Kelvin, and F$_{\lambda}$ is the flux in power per unit area per unit wavelength, $\lambda$.

\footnotetext[1]{\textsc{reflux} can be run via a web interface from \url{http://astronomy.nmsu.edu/jlcough/reflux.html}. Additional details as to how to set-up a model are presented there.}

\textsc{reflux} takes input parameters for a specified system and feeds them to the ELC program, which generates an intensity map of the system at a specified phase and wavelength, composed of N points evenly spaced on a grid around each star and the disk, if present. \textsc{reflux} then computes the system's center of light position using the formula \begin{equation} (X,Y) = \sum_{i=0}^{N}F_{i}\cdot(x,y)_{i}, \end{equation} where ($X,Y$) is the system's center of light (``photocenter''), with the center of mass located at (0,0), and F$_{i}$ is the flux of a grid point $i$, located at ($x,y$). This is done for a complete orbit at 8 different wavelengths, currently chosen to be the standard $UBVRIJHK$ bandpasses, with the astrometric reflex motion output in $\mu$as using an estimate of the system's distance. \textsc{reflux} also calculates the observed spectral energy distribution (SED) of both the sum and the individual components (as seen at quadrature) for comparison to multi-wavelength photometry to help constrain system parameters. \textsc{reflux} will also output a 3D animated gif of the system, the apparent multi-wavelength astrometric orbit, and the $x$ and $y$ components of this motion versus orbital phase. The actual light curves over an orbit are written to a text file for comparison with existing phased-resolved photometric data.

While SIM Lite only operates in the 4500 to 9000 \AA~bandpass, we include $JHK$ photometry into the modeling process to help constrain the SEDs for systems of interest. We do this because it is often possible to detect the secondary star in the near-infrared, allowing one to better quantify its contribution at the wavelengths accessible to SIM Lite. (For consistency and possible comparison to future work in NIR astrometry, we also include the J and K-band astrometric motions in the output plots.) Thus, before one can model the expected reflex motions for an IB, it is critical to have both reasonable parameters for the system, as well as multi-color photometry. \textsc{reflux} requires as inputs estimates of the masses of the two stellar components, their temperatures, their radii, the orbital period, and orbital inclination. The eccentricity of the binary is assumed to be zero since in the majority of these systems the companion star fills its Roche lobe, indicating a close orbit with strong tidal forces. In addition, the $V$ magnitude and distance must be input for model normalization. If the system has an accretion disk, the temperature at its inner edge, the power-law index for the radial dependence of its temperature, and the radius of the inner and outer edges must be inputted. Alternatively, one can choose whether the disk follows a blackbody or free-free emission law. In the latter case, one has to specify and adjust a normalization constant for the luminosity of the disk emission to best match the observed SED. To demonstrate the use of \textsc{reflux}, we model several proto-typical IB systems below.

\section{Modeling the Reflex Motions of Interacting Binaries}

It is critical that one uses reasonable system parameters to first match the observed SED before relying on the output reflex motions. There are three main scenarios for the visual SEDs of IBs: 1) systems where one stellar component dominates the optical SED, 2) where more than one component contributes to the optical SED, and 3) disk-dominated systems. Besides the stellar components, and symmetric accretion disks, there are a number of other features that are present in IB systems such as accretion disk hot spots, accretion streams, magnetic structures, jets, and other outflows that both affect the SED {\it and} which might have appreciable astrometric signatures. We investigate some of these features below, but to completely cover all of the behavior exhibited by IBs is beyond the scope of the current investigation. In the following we perform case studies for three different IB scenarios, using several well known objects. Additionally, we examine how accretion disk or photospheric ``hotspots'' can distort the reflex motions of an IB. Finally, for each system we also examine the feasibility of observing the system with SIM Lite, providing rough estimates for the amount of time that will be required of SIM Lite to observe the system at a given precision, using the SIM Differential Astrometry Performance Estimator (DAPE) \citep{Plummer09}.

\subsection{IBs with SEDs Dominated by the Primary or Secondary Star}

There are quite a number of IBs where the primary or secondary star completely dominates the SED of the system. For these systems, determining an astrometric orbit is straightforward. We examine two cases: 1) QZ Vul, a LMXB with a black hole primary and a cool, main sequence secondary star, and 2) Cyg X-1, a high mass X-ray binary (HMXB) with an O supergiant primary, and a black hole ``secondary''.

\subsubsection{QZ Vul}

\begin{figure*}[t]
\centering
\begin{tabular}{cc}
\epsfig{file=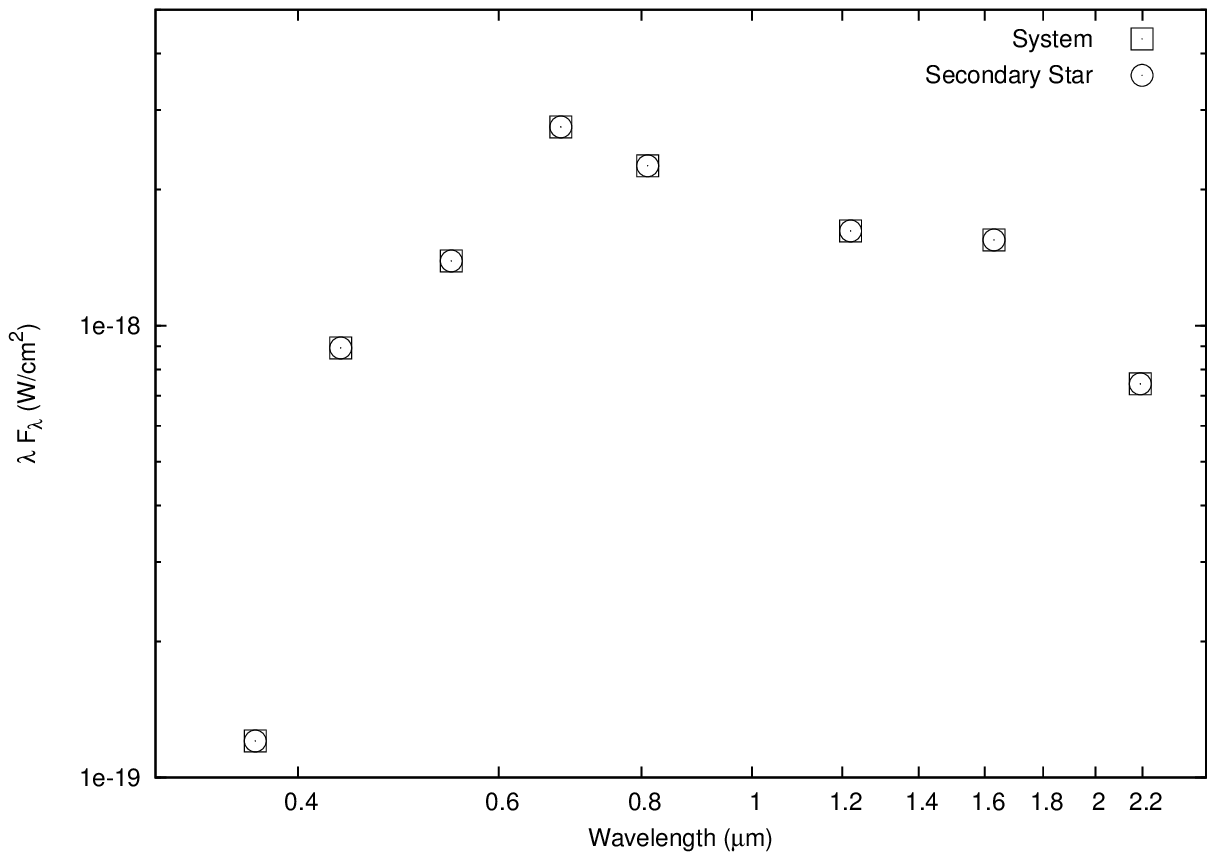, angle=0,width=0.45\linewidth} &
\includegraphics[width=0.33\linewidth]{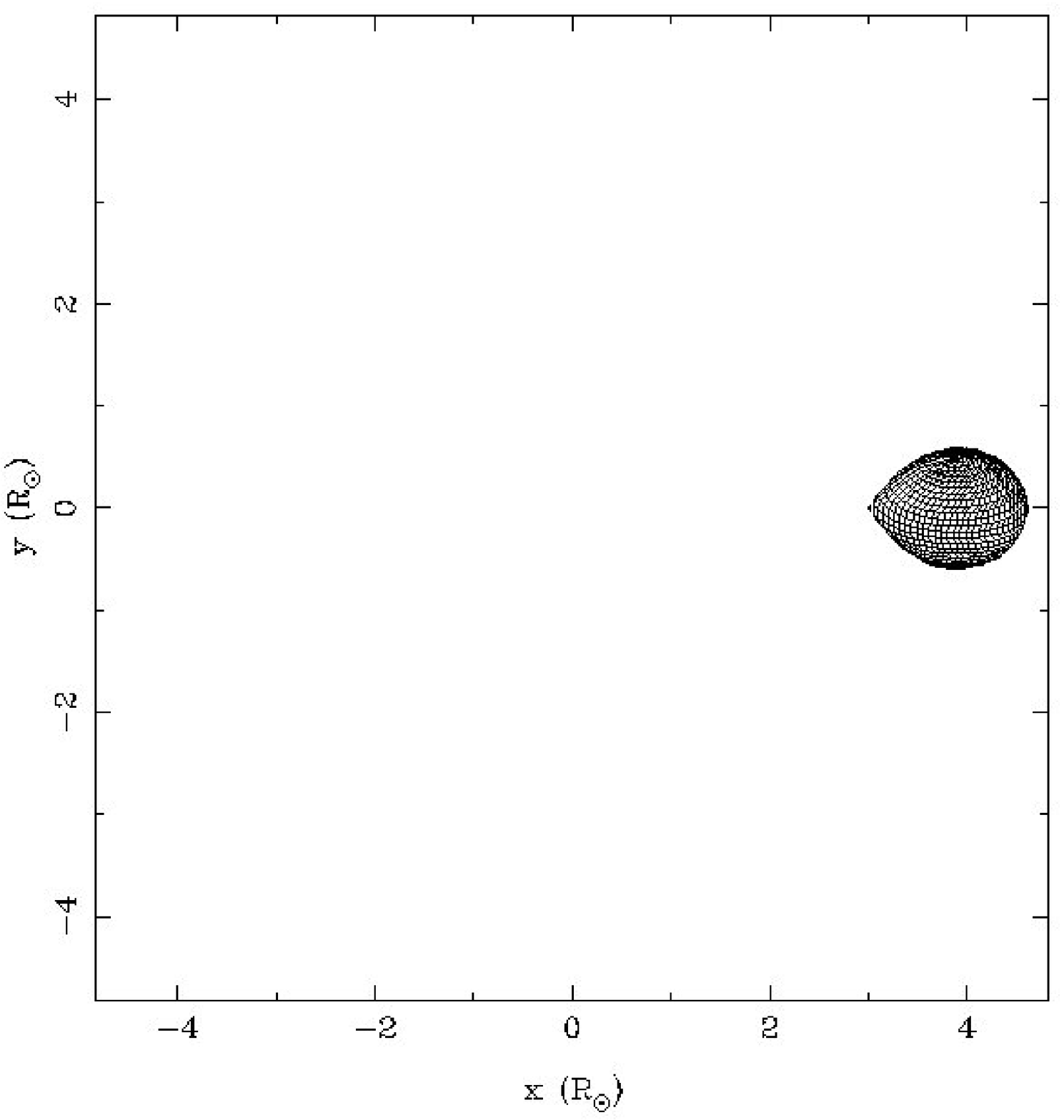} \\
\epsfig{file=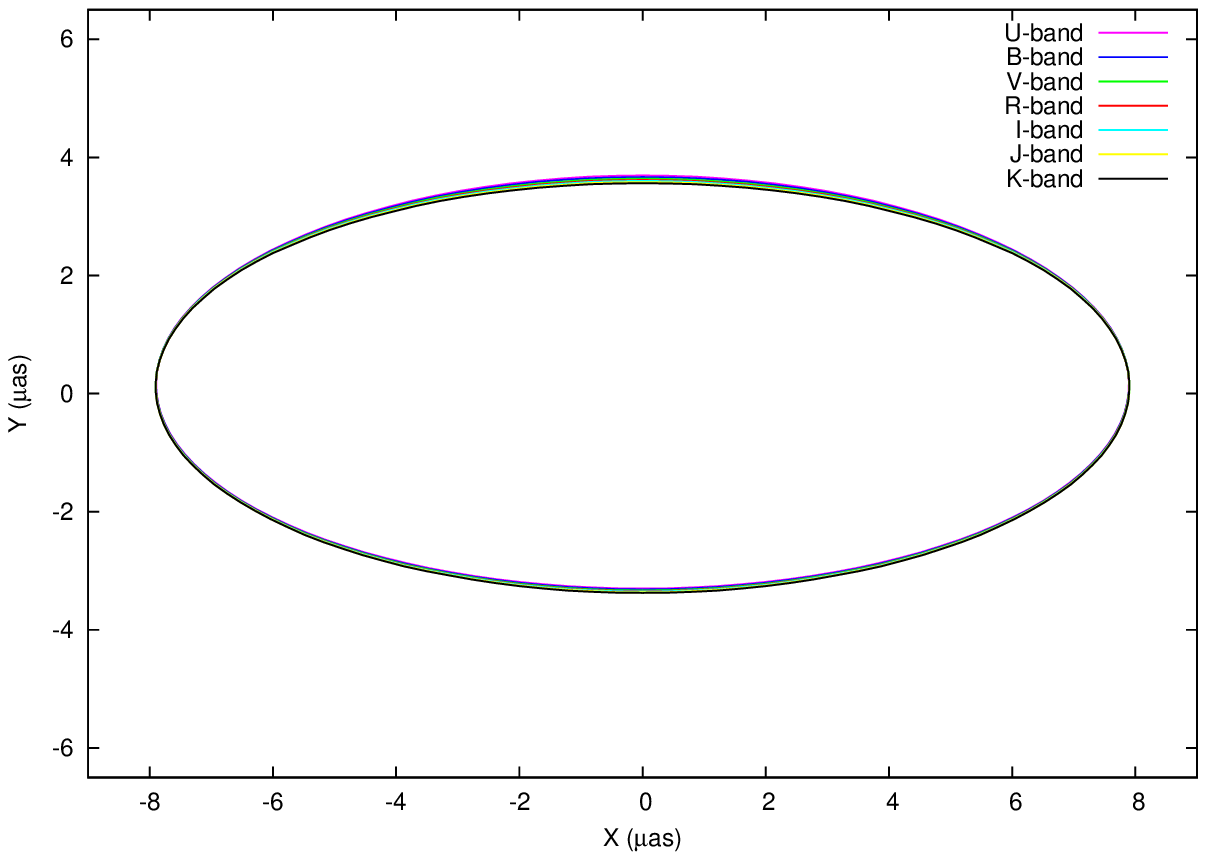, angle=0,width=0.45\linewidth} &
\epsfig{file=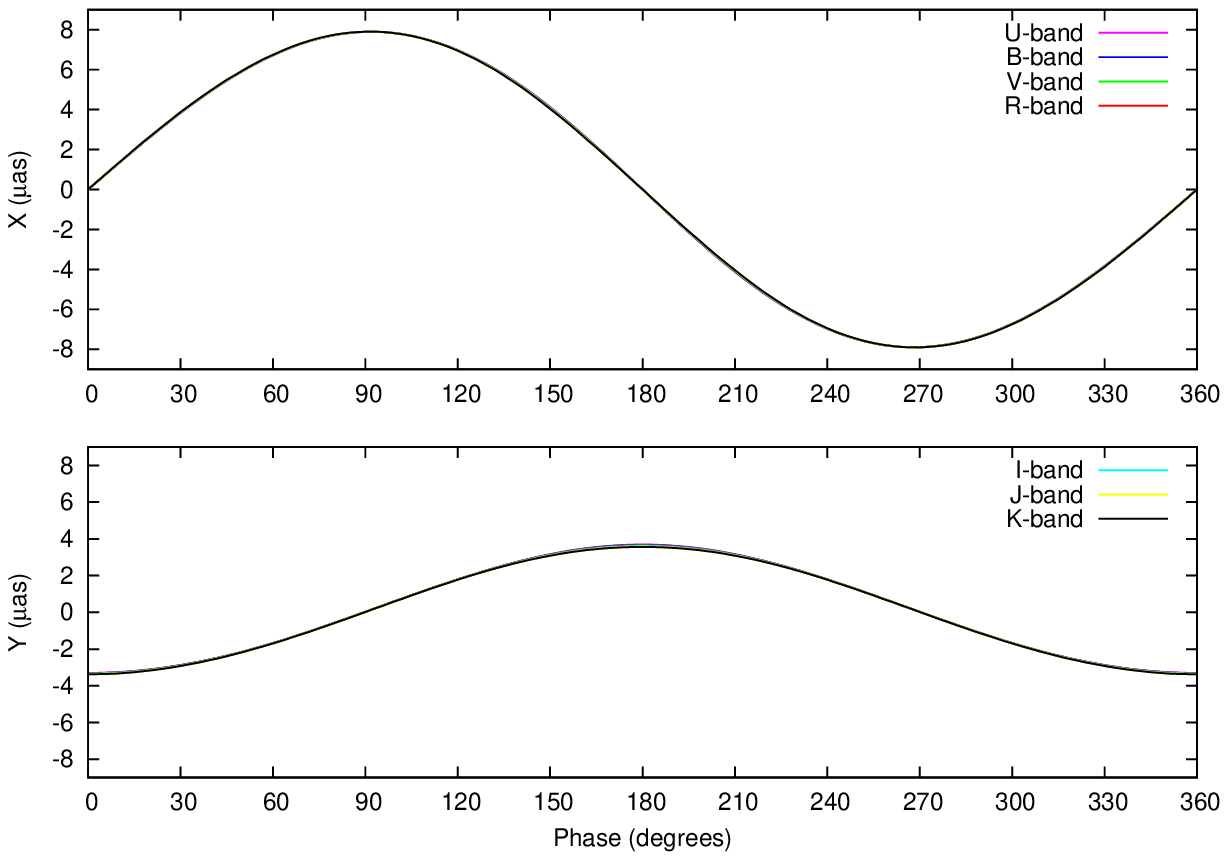, angle=0,width=0.45\linewidth} \\
\end{tabular}
\caption{\emph{Top Left:} SED plot for QZ Vul. \emph{Top Right:} 3D model of QZ Vul at Phase=90$\degr$. \emph{Bottom Left:} Reflex orbit for QZ Vul. \emph{Bottom Right:} X and Y components of the reflex orbit versus phase for QZ Vul.}
\label{qzvulmainfig}
\end{figure*}

As shown in \citet{Gelino09}, the SED of QZ Vul appears to be that of a reddened K2 dwarf from the optical through the mid-infrared. The system parameters are shown in Table~\ref{qzvultable}. These parameters were input into \textsc{reflux} to model the astrometric motions, with a \textsc{nextgen} model atmosphere used for the secondary star. The model SED, a 3D model of the system, and its wavelength-dependent reflex motions are shown in Fig.~\ref{qzvulmainfig}. Since there is only a single visible component, QZ Vul does not exhibit any discernible wavelength dependency to its reflex motions. However, since the black hole is $\sim$24 times more massive than the secondary star, the K2 dwarf has a large apparent motion, and thus even at a distance of 2.29 kpc, the system's astrometric reflex motion ($\sim$8 $\mu$as), in theory, would be detectable with microarcsecond astrometry. However, in practice, this particular system is so faint (V = 21.2), and has such a short period (0.3342 days), that, according to DAPE, SIM Lite can not reach the needed precision without integrating for longer than the orbital period. Thus, one would need to find a much closer, and thus brighter, LMXB to observe with SIM Lite. The estimated orbital inclination for QZ Vul is 64$^{\circ}$ \citep{Gel01}, and the only effect one would see with a QZ Vul type system with a different inclination is a reduction or amplification of the y-component of the astrometric motion, corresponding to an increase or decrease of the inclination respectively.

\begin{deluxetable}{lc}
\tablewidth{0pt}
\tablecaption{Parameters for the QZ Vul System}
\tablecolumns{2}
\tablehead{Parameter & Value\tablenotemark{a}}
\startdata
Magnitude (V) & 21.2\\
Distance (kpc) & 2.29\\
Inclination ($\degr$) & 64.0\\
Period (Days) & 0.3342\\
Eccentricity & 0.0\\
Mass of Star 1 (M$_{\sun}$) & 0.32\\
Mass of Star 2 (M$_{\sun}$) & 7.7\\
Radius of Star 1 (R$_{\sun}$) & 0.397\tablenotemark{b}\\
Radius of Star 2 (R$_{\sun}$) & 0.0\\
T$_{\rm eff}$ of Star 1 (K) & 4500\\
T$_{\rm eff}$ of Star 2 (K) & 0
\enddata
\label{qzvultable}
\tablenotetext{a}{Values from Gelino (2001)}
\tablenotetext{b}{Secondary star fills Roche lobe}
\end{deluxetable}

\subsubsection{Cyg X-1}

\begin{figure*}[t]
\centering
\begin{tabular}{cc}
\epsfig{file=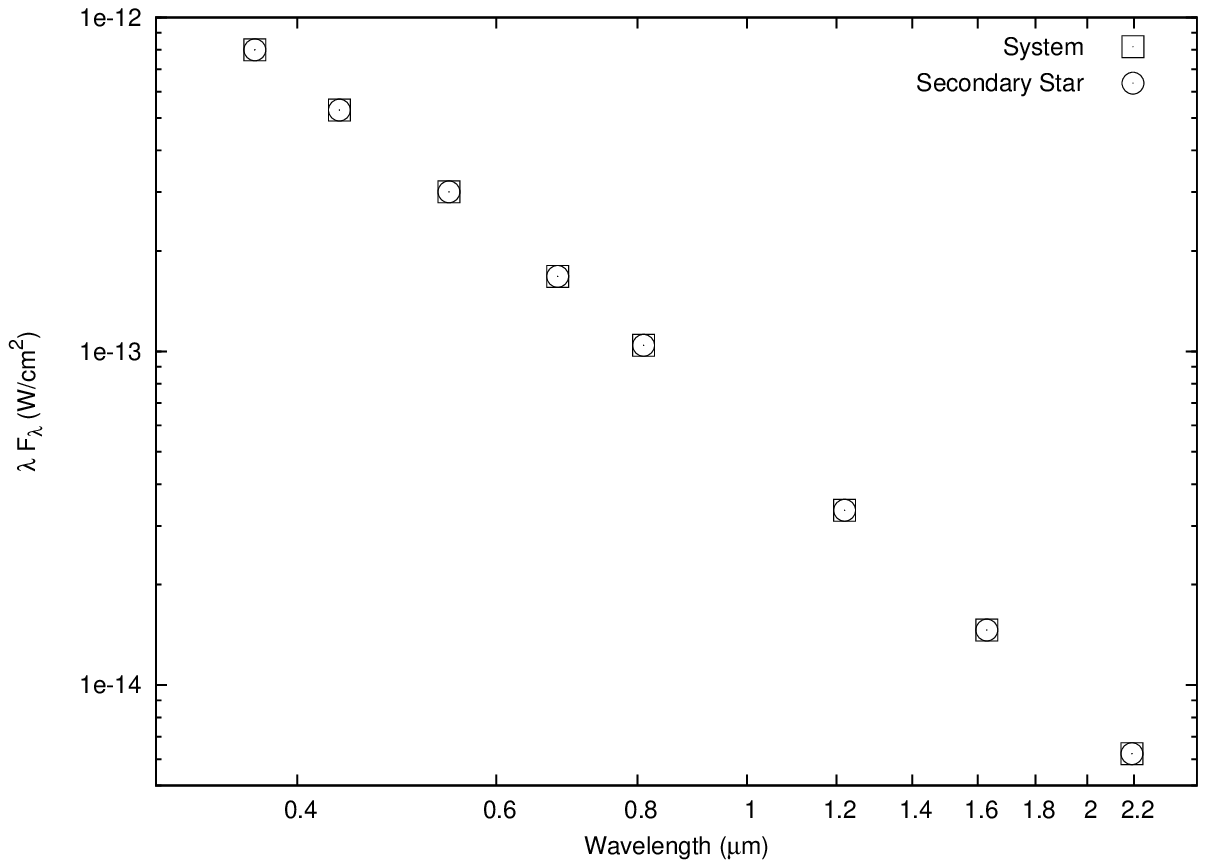, angle=0,width=0.45\linewidth} &
\includegraphics[width=0.33\linewidth]{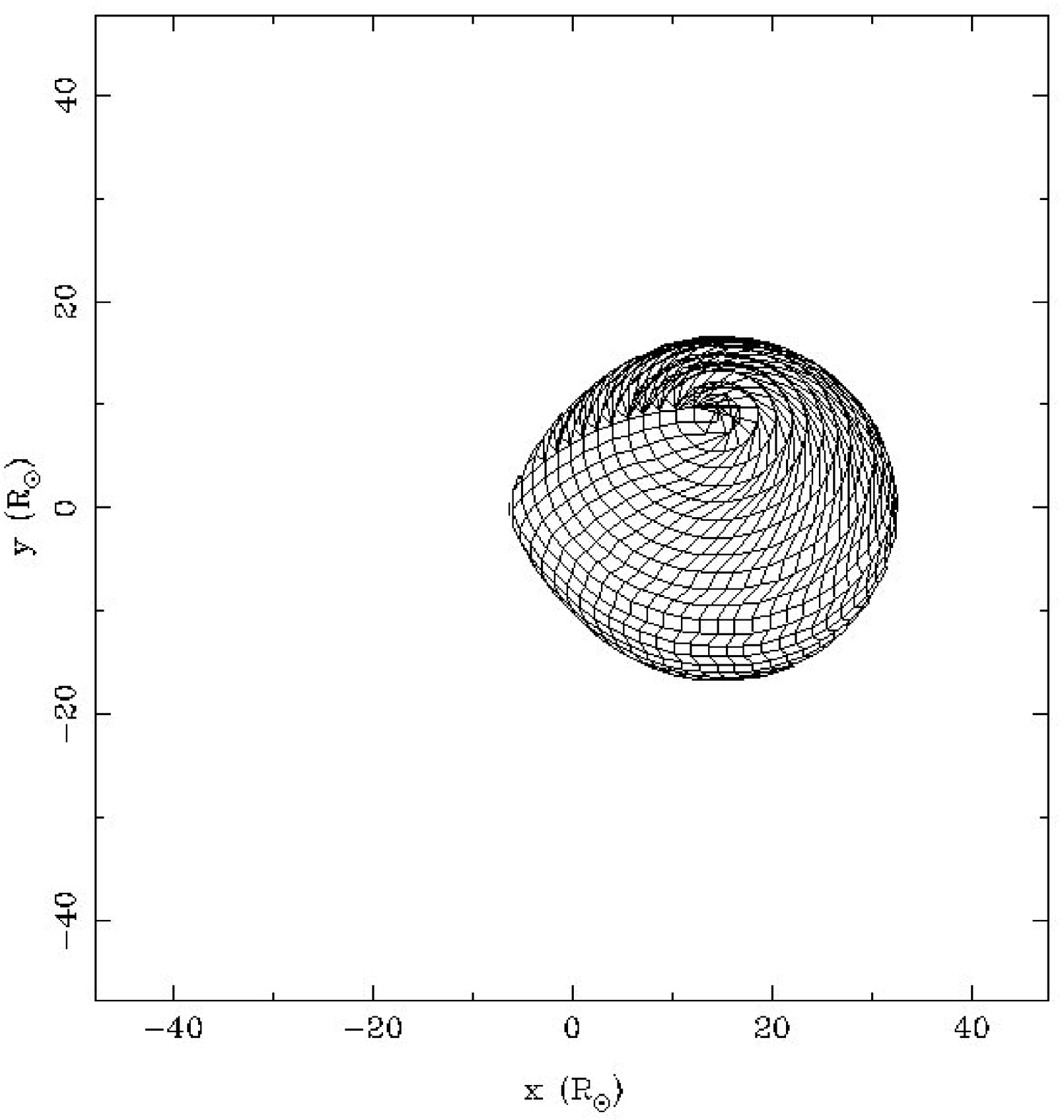} \\
\epsfig{file=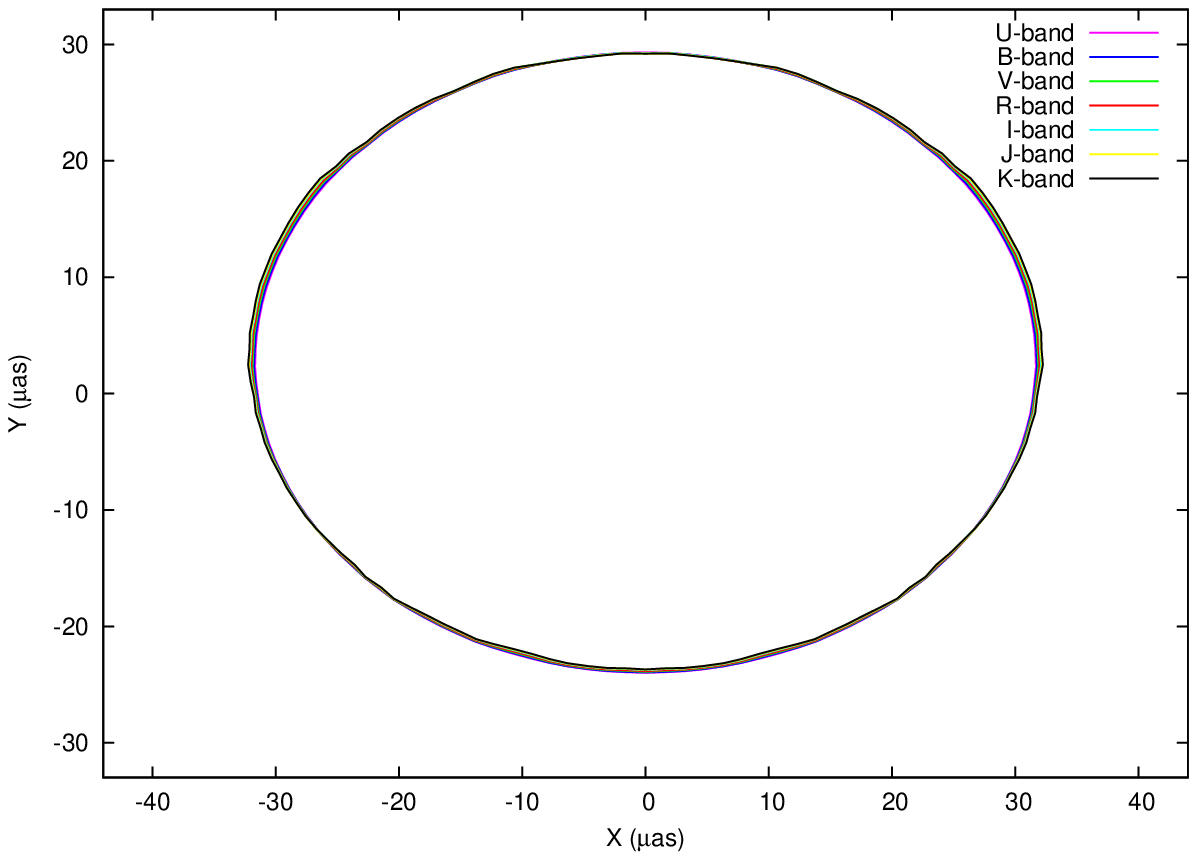, angle=0,width=0.45\linewidth} &
\epsfig{file=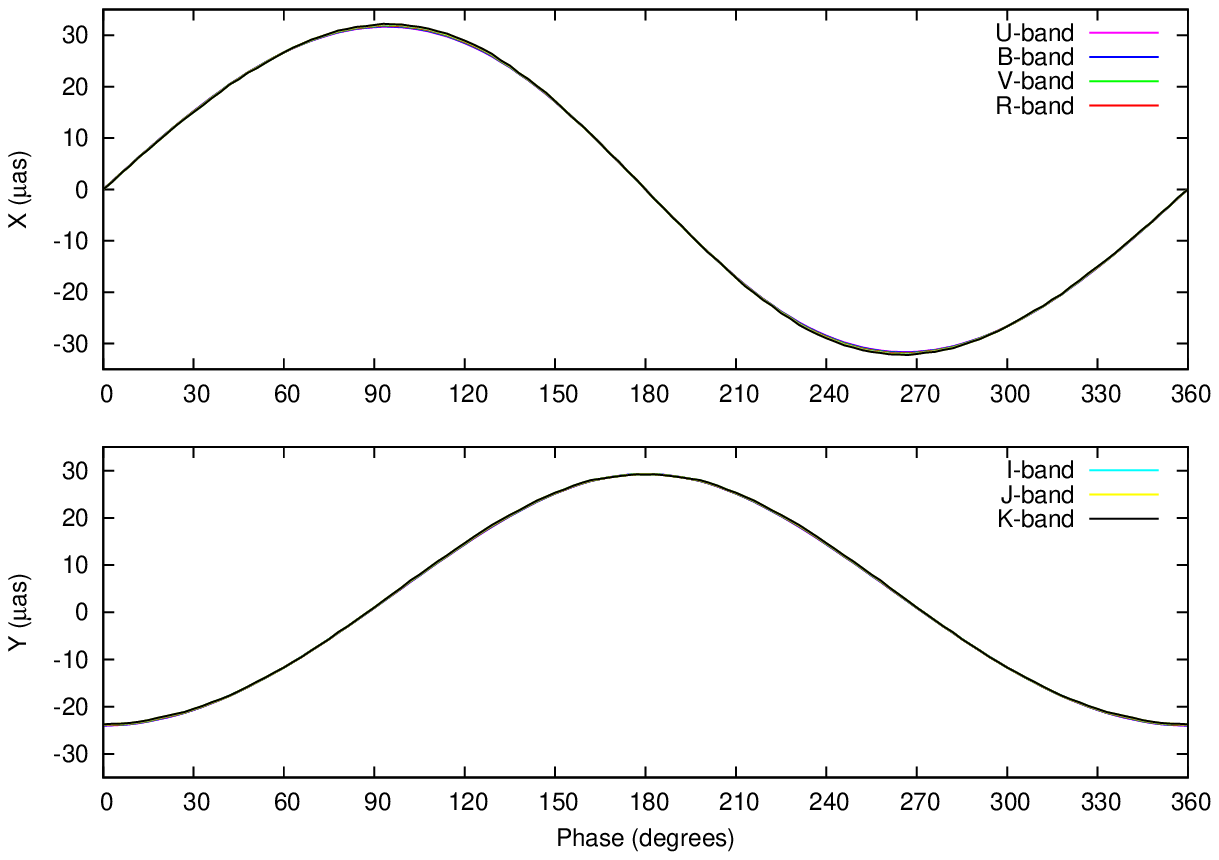, angle=0,width=0.45\linewidth} \\
\end{tabular}
\caption{\emph{Top Left:} SED plot for Cyg X-1. \emph{Top Right:} 3D model of Cyg X-1 at Phase=90$\degr$. \emph{Bottom Left:} Reflex orbit for Cyg X-1. \emph{Bottom Right:} X and Y components of the reflex orbit versus phase for Cyg X-1.}
\label{cygx1mainfig}
\end{figure*}

Cyg X-1 is a well-known HMXB dominated by an O-type supergiant orbited by a 10 M$_{\sun}$ black hole with a period of 5.6 d. Cyg X-1 exhibits a wide variety of behavior, such as highly variable X-ray and radio emission, some of which is presumably due to a relativistic jet. The optical light also varies, but much more weakly ($\Delta$m $\approx$ 0.05 mag). Thus, in contrast to QZ Vul, the primary star will be the visible source in this system. We model this system using the parameters listed in Table~\ref{cygx1table}, using a blackbody for the O star. As can be seen in Fig.~\ref{cygx1mainfig}, there is no wavelength dependence to the astrometric reflex motion, and a single bandpass is sufficient to determine the orbit. Even at a distance of 2.1 kpc, the large separation of the components in Cyg X-1 produces a significant astrometric signature ($\sim$25 $\mu$as), even though it is the more massive component that is responsible for the visible astrometric motion in this system. Given the brightness of the system (V = 8.95), and the long orbital period, this system is easy to observe with SIM Lite. According to DAPE, 10 individual measurements, each with 2.5 $\mu$as precision, could be obtained in only a total of 1 hour and 40 minutes of mission time, (given 5 minutes of target integration time per visit, broken into 5, 1-minute chops between the target and reference star.) This would provide more than sufficient high-precision measurements to obtain a good solution for this system, directly yielding the inclination of the system and the semi-major axis of the primary star's orbit, and thus indirectly the masses of the components. For systems in which one component completely dominates the visual SED, having accurate photometry for the system is unnecessary in interpreting the astrometric data.

\begin{deluxetable}{lc}
\tablewidth{0pt}
\tablecaption{Parameters for the Cyg X-1 System}
\tablecolumns{2}
\tablehead{Parameter & Value\tablenotemark{a}}
\startdata
Magnitude (V) & 8.95\\
Distance (kpc) & 2.10\\
Inclination ($\degr$) & 35\\
Period (Days) & 5.566\\
Eccentricity & 0.0\\
Mass of Star 1 (M$_{\sun}$) & 17.8\\
Mass of Star 2 (M$_{\sun}$) & 10.1\\
Radius of Star 1 (R$_{\sun}$) & 16\\
Radius of Star 2 (R$_{\sun}$) & 0.0\\
T$_{\rm eff}$ of Star 1 (K) & 32000\\
T$_{\rm eff}$ of Star 2 (K) & 0
\enddata
\label{cygx1table}
\tablenotetext{a}{Values from \citet{Herrero95}}
\end{deluxetable}

\subsection{IBs Where the SED is Not Fully Dominated by Either the Primary or Secondary Star}

\subsubsection{SS Cygni}

SS Cyg is a bright, well-known, cataclysmic variable. In quiescence, SS Cyg has an apparent magnitude of $V$ = 12.2. During outbursts, it brightens to $V$ $\sim$ 8.8. It has an orbital period of 6.603 hrs, and the secondary star has a spectral type of K5 \citep{Harr04}. \citet{Dubus04} obtained near-simultaneous multi-wavelength photometry of SS Cyg. As shown in Figure 5 of \citet{Harr07} the white dwarf and accretion disk dominate the blue end of the SED, while the secondary star becomes prominent in the red and near-infrared. \citet{Harr07} modeled the SED of SS Cyg as a combination of the two stellar components plus a free-free accretion disk around the white dwarf. Due to the availability of simultaneous $UBVRIJHK$ photometry, and well-constrained stellar parameters, SS Cyg is an ideal system to investigate what happens when more than one component in the binary system is visible. As we demonstrate, SS Cyg has strongly wavelength-dependent astrometric reflex motions. We only consider the reflex motion in quiescence, since during outburst, the luminosity of the system is completely dominated by the hot (10,000~K), optically thick accretion disk.

\begin{figure*}[t]
\centering
\begin{tabular}{cc}
\epsfig{file=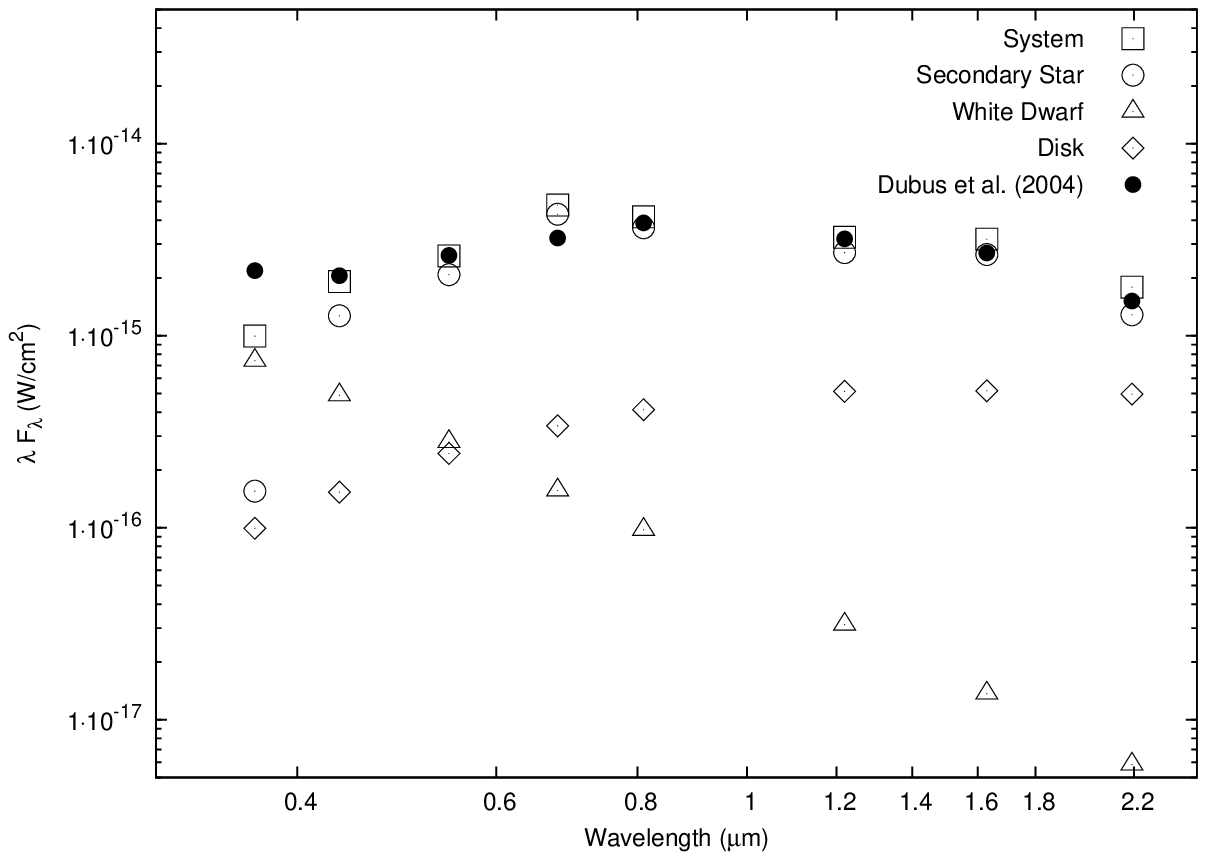, angle=0,width=0.45\linewidth} &
\includegraphics[width=0.33\linewidth]{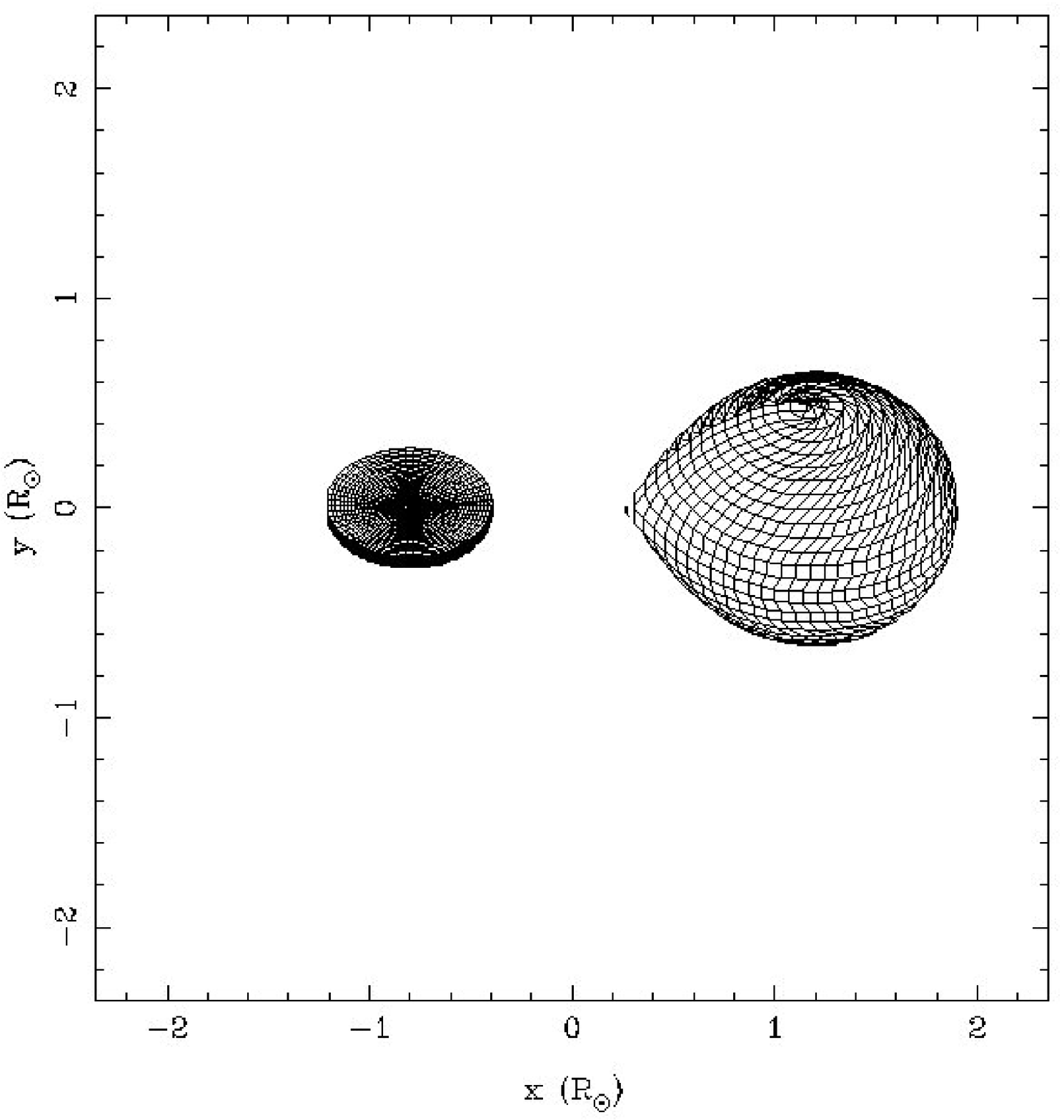} \\
\epsfig{file=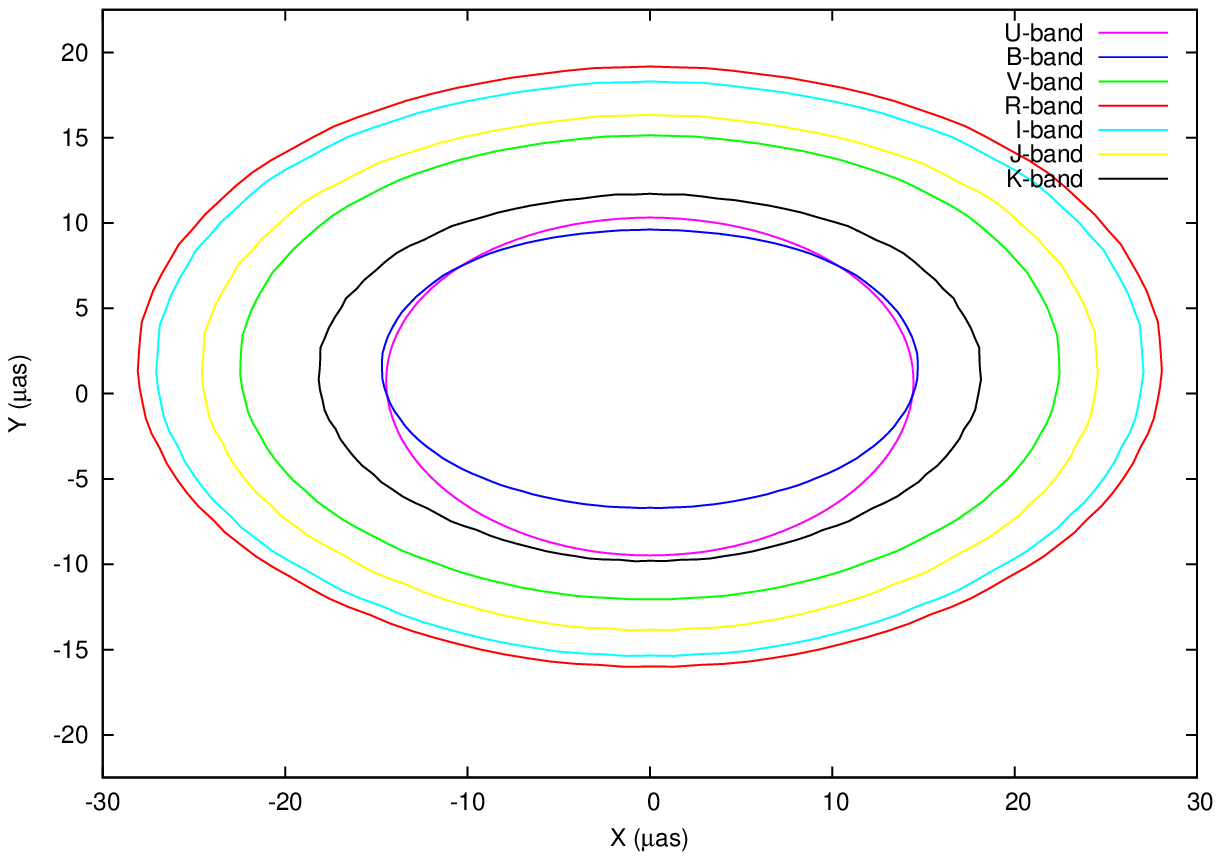, angle=0,width=0.45\linewidth} &
\epsfig{file=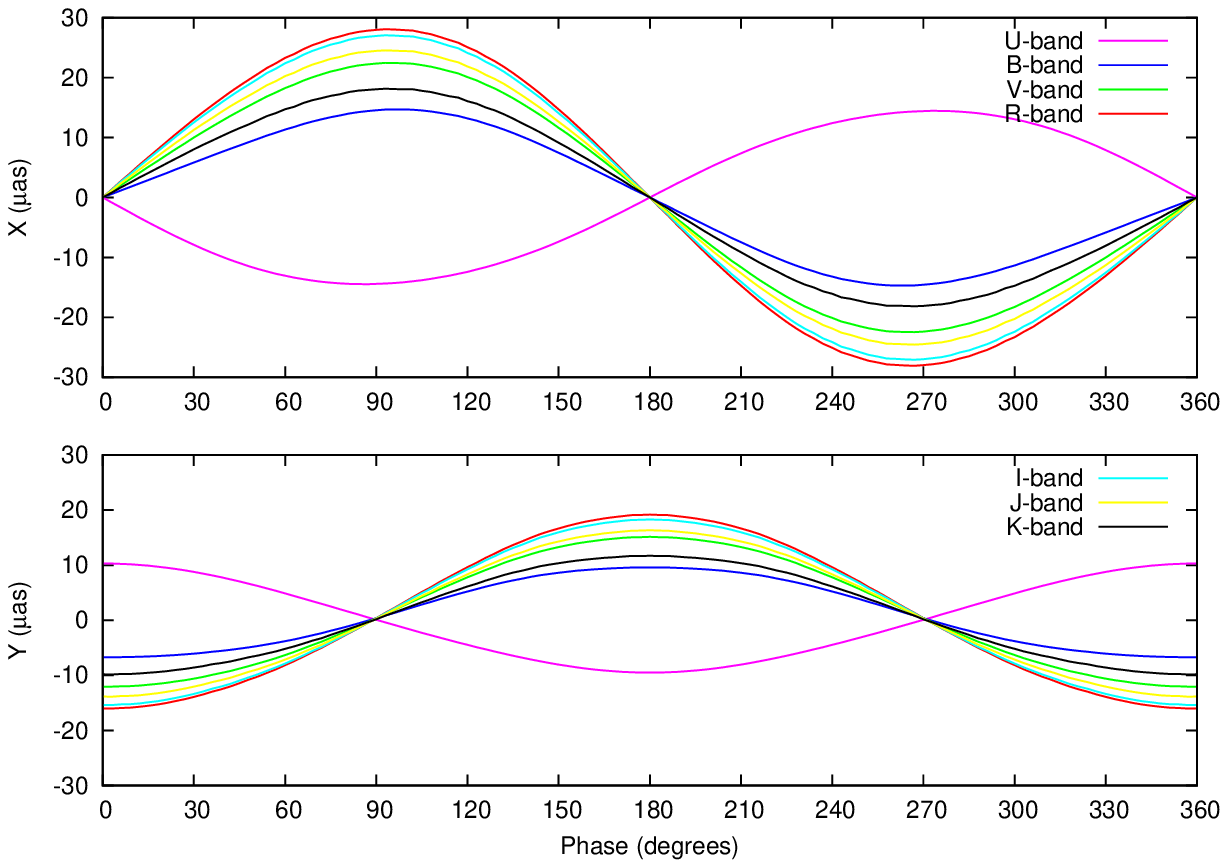, angle=0,width=0.45\linewidth} \\
\end{tabular}
\caption{\emph{Top Left:} SED plot for SS Cyg compared to photometry from \citet{Dubus04}. \emph{Top Right:} 3D model of SS Cyg at Phase=90$\degr$. \emph{Bottom Left:} Reflex orbit for SS Cyg. \emph{Bottom Right:} X and Y components of the reflex orbit versus phase for SS Cyg.}
\label{sscygmainfig}
\end{figure*}

We have generated a model system with the parameters listed in Table~\ref{sscygtable}, where the temperatures and radii of the stellar components have been derived from the literature. In the case of SS Cyg, we used a \textsc{nextgen} model atmosphere for the secondary star, and assumed a blackbody for the white dwarf primary. We attempted both blackbody and free-free accretion disk models, and found that the free-free model provides the best match to the photometry. The results are shown in Fig.~\ref{sscygmainfig}, where the model SED is compared to the photometry, with the contributions from each of the three components shown with separate symbols. As can be seen from these plots, the white dwarf dominates the flux contribution in the $U$-band, and fades thereafter, with the secondary dominating from $B$ through $K$ (though the accretion disk contribution becomes quite important in the infrared). A wire grid representation of the SS Cyg system at phase 0.25 is also shown in Fig.~\ref{sscygmainfig}. As shown in the bottom-left panel, the amplitudes of the reflex motions smoothly increase from $B$ (semi-major axis $a$ = 12 $\mu$as) to $R$ ($a$ = 28 $\mu$as), and then decline at longer wavelengths due to the contribution of the accretion disk. The reflex motion in the $U$-band is larger than in $B$, as it is almost completely dominated by the WD primary. In the $B$-band, the two stars have similar luminosities, and thus the motions are a mixture of the individual astrometric orbits. If there was no accretion disk, and the white dwarf was invisible, the total amplitude of the reflex motions for the secondary star in this system would be $a$ = 34.1 $\mu$as. Thus, the true reflex motions are diluted, and one cannot determine the actual astrometric orbit without modeling the system. However, with proper modeling, one can deconvolve the motion of the white dwarf + accretion disk and the secondary star, thus directly yielding the semi-major axis of each component's orbit, which when combined with the well-known period and distance to the system, directly yields absolute masses for each component, (see \S4).

\begin{deluxetable}{lc}
\tablewidth{0pt}
\tablecaption{Parameters for the SS Cyg System}
\tablecolumns{2}
\tablehead{Parameter & Value\tablenotemark{a}}
\startdata
Magnitude (V) & 8.8 (max) 12.20 (min)\\
Distance (pc) & 159.5\\
Inclination ($\degr$) & 50.5\\
Period (Days) & 0.275130\\
Eccentricity & 0.0\\
Mass of Star 1 (M$_{\sun}$) & 0.555\\
Mass of Star 2 (M$_{\sun}$) & 0.812\\
Radius of Star 1 (R$_{\sun}$) & 0.684\tablenotemark{b}\\
Radius of Star 2 (R$_{\sun}$) & 0.015\\
T$_{\rm eff}$ of Star 1 (K) & 4400\\
T$_{\rm eff}$ of Star 2 (K) & 35000\\
Disk Inner Radius (R$_{\sun}$) & 0.022\\
Disk Outer Radius (R$_{\sun}$) & 0.407\\
Disk Inner Temperature (K) & 10000\tablenotemark{c}\\
Disk Temp Power-Law Exponent & 0.0
\enddata
\tablenotetext{a}{Values from \citet{Rob09}}
\tablenotetext{b}{Star fills its Roche Lobe}
\tablenotetext{c}{Disk is free-free}
\label{sscygtable}
\end{deluxetable}

Careful examination of the wavelength-dependent astrometric orbits reveals that the actual center about which the motion occurs is wavelength dependent. Note that the $U$-band reflex motion is centered on (0,0), but at the other wavelengths, the motions are offset from this position. This subtle effect has two causes: 1) the secondary star in SS Cyg fills its Roche lobe, and therefore has a teardrop shape, and 2) gravity brightening for non-degenerate stars is significant. These two effects combine to offset
the center of light of the secondary star from its center of mass. This creates an offset in the astrometric motions in the bandpasses where the secondary star dominates the systemic luminosity. Since the WD is spherical, and does not have gravity brightening effects, the $U$-band motion remains centered at (0,0). For more detailed modeling on the wavelength dependence of the astrometric center due to gravity brightening, see \citet{Coughlin10}.

One of the important consequences of these effects is that the actual orbital inclination is harder to derive. Naively, one would assume that the orbital inclination can be calculated from the ratio of the minor and major axes of the orbital ellipse. If we do this in the $R$-band [$i$ = cos$^{\rm -1}$(17.5/28.0)] we derive an orbital inclination of 51.4$^{\circ}$, instead of the input value of 50.5$^{\circ}$. Depending on the bandpass, the orbital inclinations derived for our simulation of SS Cyg range from a minimum of 47$^{\circ}$ in the $U$-band, to a maximum of 51.4$^{\circ}$ in the $R$-band! While it is likely that these differences will be lost in the astrometric noise for most IB systems, there will be a small number of IBs (and many non-interacting binary systems) where this difference should be detectable, and proper modeling is required to accurately extract the system's inclination.

With respect to observing this system with SIM Lite, according to DAPE, even in its low state with V = 12.1, a total mission time of 6 hours and 40 minutes would provide 10 measurements with respect to an astrometric standard star, each with 1.7 $\mu$as precision and an individual integration time of 20 minutes per measurement, each composed of 20 chops between the target and reference stars, with 1 minute exposures on the target and 30 second exposures on the reference. As a minimum of 7 points are needed to solve for a full astrometric orbit, (given that 7 parameters describe an orbit), and as the orbital period of SS Cyg is $\sim$6.6 hours, this would successfully measure the full astrometric orbit with limited orbital smearing, (see \S\ref{solvesec}).

SS Cyg can be used as a template for cases of CVs and LMXBs where the secondary star is clearly visible in the optical. It simply requires changing the input values of the stellar components, the relative prominence (and nature) of the accretion disk, and altering the orbital inclination. Other changes, such as orbital period, mass ratio, and distance, act to scale the amplitude of the reflex motions. However, it remains critically important to have simultaneous multi-wavelength photometry, and a reasonable handle on the system properties to derive accurate astrometric orbits for these types of objects. As one would expect, the wavelength dependent behavior of the astrometric motions for these types of systems is more complex than for systems with only a single, visible component, but the potential scientific payoff is much higher. When varying the inclination angle for a SS Cyg type system, the y component of the astrometric motion decreases with increasing inclination angle for all wavelengths as expected, but at large inclination angles eclipse effects produce significant deviations from simple sinusoidal reflex motions.

\subsection{IBs with Accretion Disk Dominated SEDs}

The majority of LMXBs, and many CVs, have their optical light dominated by the accretion disk. The accretion disk surrounds the compact object, and thus the observed astrometric motion is that of the compact object, although there are some complications arising from the extended structure of the disk. Here we examine two cases, the short period CV system V592 Cas, and the well-known LMXB Sco X-1.

\subsubsection{V592 Cas}

\begin{figure*}[t]
\centering
\begin{tabular}{cc}
\epsfig{file=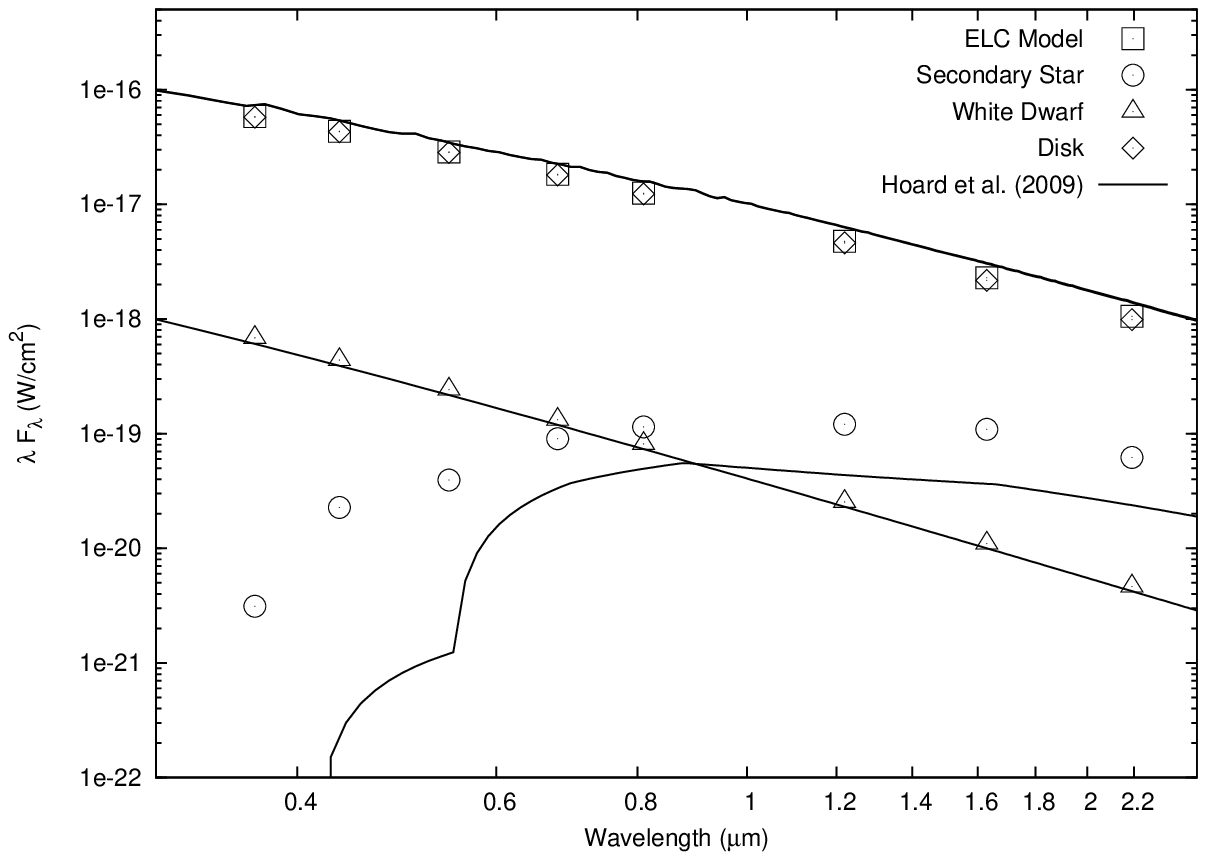, angle=0,width=0.45\linewidth} &
\includegraphics[width=0.33\linewidth]{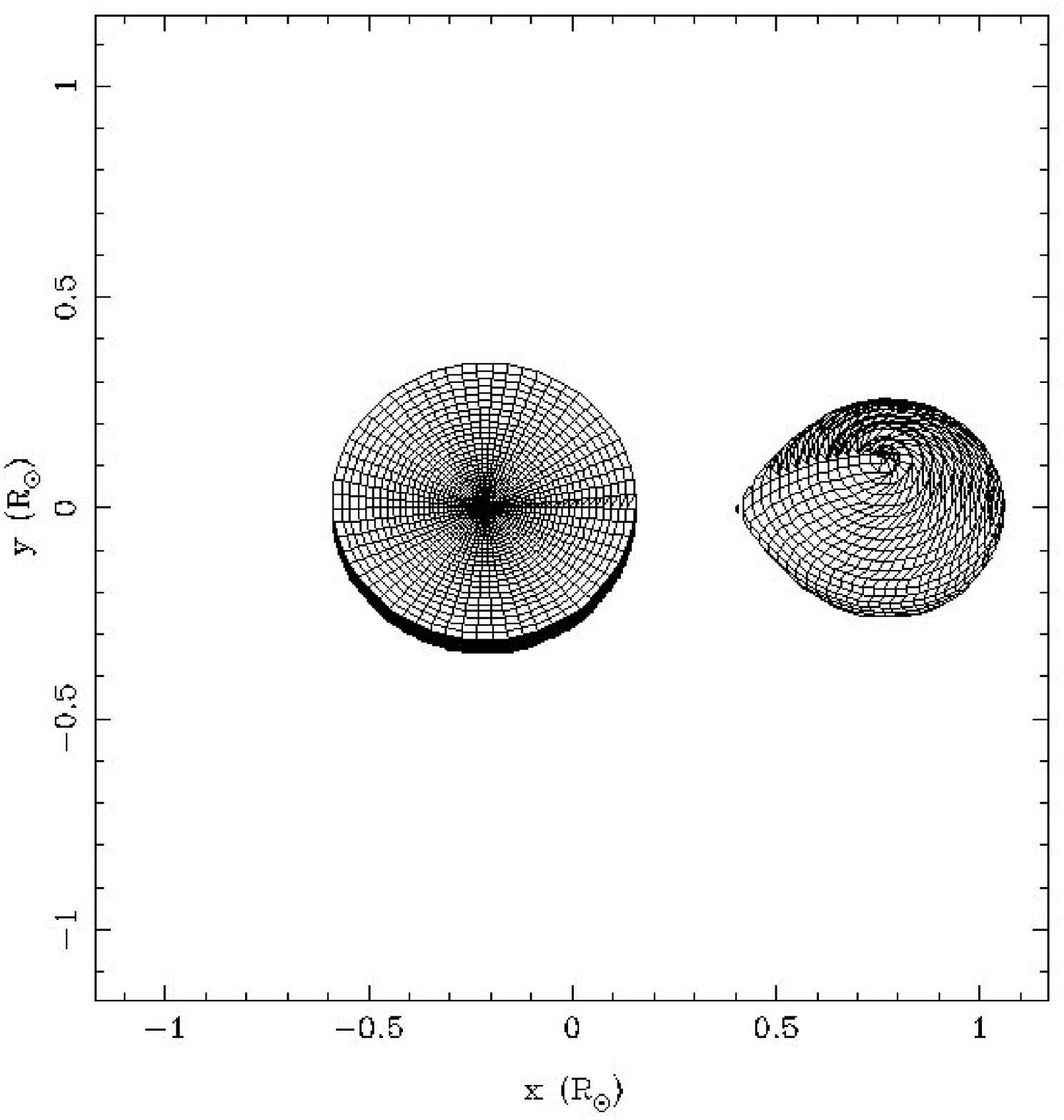} \\
\epsfig{file=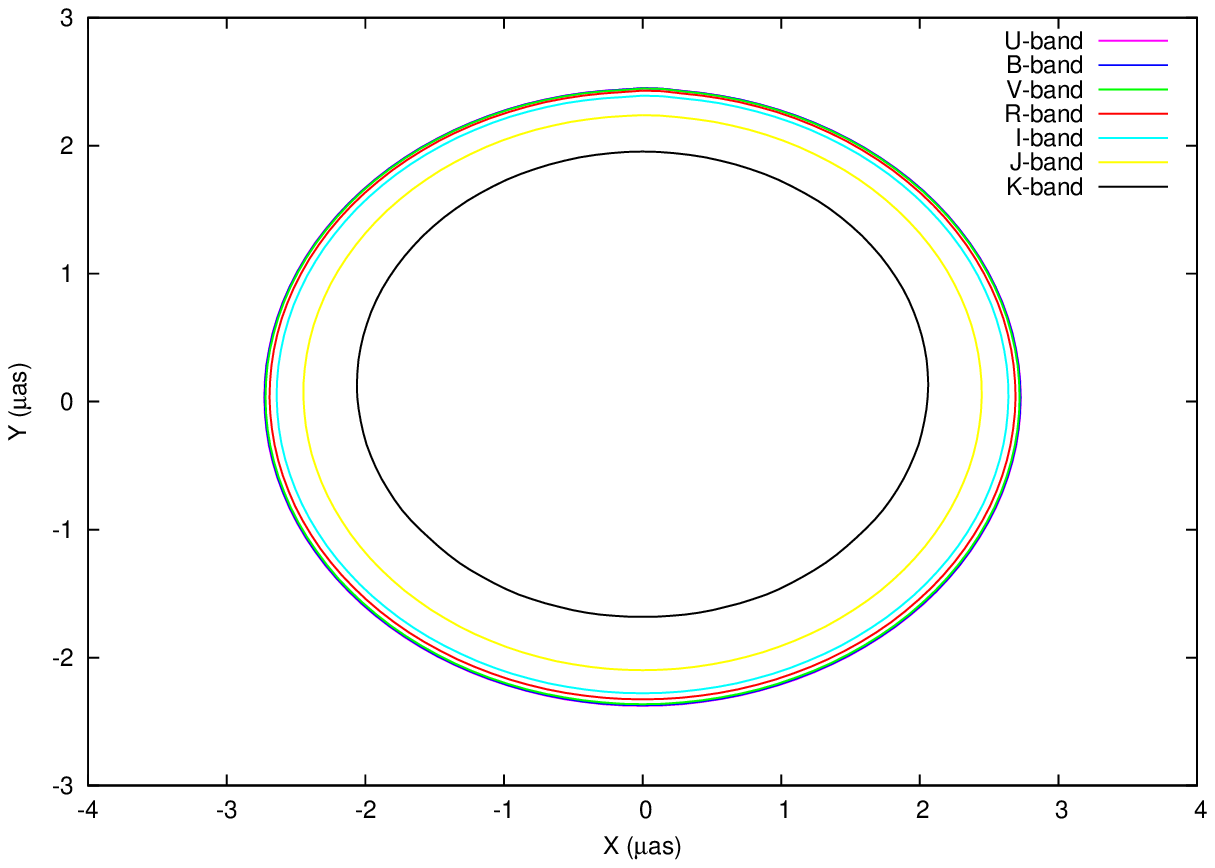, angle=0,width=0.45\linewidth} &
\epsfig{file=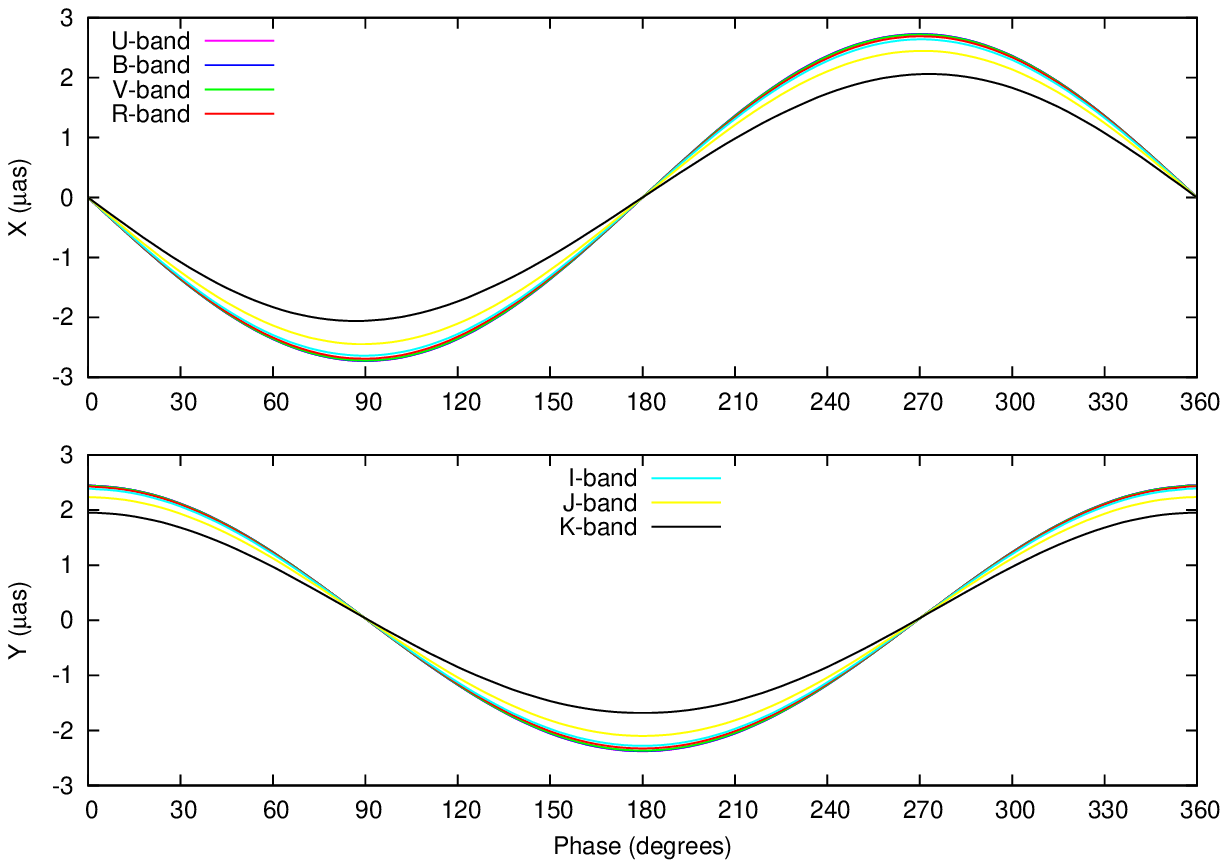, angle=0,width=0.45\linewidth} \\
\end{tabular}
\caption{\emph{Top Left:} SED plot for V592 Cas compared to models from \citet{Hoard09}. \emph{Top Right:} 3D model of V592 Cas at phase 0.5. \emph{Bottom Left:} Reflex orbit for V592 Cas. \emph{Bottom Right:} X and Y components of the reflex orbit versus phase for V592 Cas.}
\label{v592casmainfig}
\end{figure*}

V592 Cas is a short-period (P$_{\rm orb}$ = 2.76 hr) CV  system consisting of a Roche lobe-filling M dwarf, a very hot white dwarf, and a disk that dominates the luminosity at optical and near-infrared wavelengths. The majority of short period CVs closely resemble V592 Cas, and thus it can act as a prototype for disk-dominated CVs. The SED of the system has been investigated by \citet{Hoard09}, who reproduced the observed photometry using a model that included a red dwarf, white dwarf, and a two component blackbody disk around the white dwarf consisting of an inner flat component and an outer flared component. To model the mid-infrared excess, a circumbinary dust disk was also included. This dust disk is only important at mid-infrared wavelengths, and thus can be ignored in this present study. We model the reflex motions of V592 Cas using the parameters from \citet{Hoard09} shown in Table~\ref{v592castable}, while using a single component blackbody disk model with a flare inclination that is the average of the angles in their two component disk model. The SED produced by \textsc{reflux} is compared to that of \citet{Hoard09} in Fig. \ref{v592casmainfig}, where the contributions from each component are also shown. We find a greater contribution of flux from the M dwarf compared to \citet{Hoard09}, by a factor of $\sim$1.75. This is most likely due to the fact that \citet{Hoard09} used the spectrum of a field M5.0V dwarf, while our model, which incorporates full Roche geometry, produces a star with the same effective temperature, but a larger surface area by the same factor of $\sim$1.75, adopting the radius for an isolated M5.0V as 0.21 R$_{sun}$, as shown by recent measurements of low-mass stars \citep[c.f.,][]{Lopez-Morales07}. Either way, the M dwarf is a few percent of the total system flux, and thus does not greatly affect the derived astrometric measurements.

As seen in Figure~\ref{v592casmainfig}, due to the domination by the disk, the system shows less wavelength dependent motion than SS Cyg. However, there is still a discernible effect towards longer wavelengths as the secondary becomes a more significant source of flux. Thus, with enough measurements, one could in theory derive the individual component masses. The influence of the orbital inclination is readily apparent, and is uncontaminated by the secondary star gravity effects that were visible in our models for SS Cyg. With respect to SIM Lite observing time, this system would require a moderate amount of SIM Lite mission time, given its small amplitude and short period. According to DAPE, if we constrain individual integration times to 15 minutes, (0.1 of the orbital period), to prevent orbital smearing, one could obtain 2.3 $\mu$as precision per visit (composed of 20 target-reference chops of 45 second target integration each), which is on par with the amplitude of the observed reflex motions. Thus, in comparison to SS Cyg and Cyg X-1, discussed above, one might need more than 10 measurements to build up signal-to-noise and really nail down the system parameters. 30 measurements would take a total mission time of 17.5 hours, and thus we can say with reasonable confidence that this system would require less than a day of total mission time to observe so that the astrometric parameters could be obtained with reasonable precision.

\begin{deluxetable}{lc}
\tablewidth{0pt}
\tablecaption{Parameters for the V592 Cas System}
\tablecolumns{2}
\tablehead{Parameter & Value\tablenotemark{a}}
\startdata
Magnitude (V) & 12.8\\
Distance (pc) & 364.0\\
Inclination ($\degr$) & 28.0\\
Period (Days) & 0.115063\\
Eccentricity & 0.0\\
Mass of Star 1 (M$_{\sun}$) & 0.210\\
Mass of Star 2 (M$_{\sun}$) & 0.751\\
Radius of Star 1 (R$_{\sun}$) & 0.270\tablenotemark{b}\\
Radius of Star 2 (R$_{\sun}$) & 0.014\\
T$_{\rm eff}$ of Star 1 (K) & 3030\\
T$_{\rm eff}$ of Star 2 (K) & 45000\\
Disk Inner Radius (R$_{\sun}$) & 0.0106\\
Disk Outer Radius (R$_{\sun}$) & 0.371\\
Disk Inner Temperature (K) & 109,700\tablenotemark{c}\\
Disk Temp Power-Law Exponent & $-$0.75
\enddata
\tablenotetext{a}{Values from \citet{Hoard09}}
\tablenotetext{b}{Star fills its Roche Lobe}
\tablenotetext{c}{Disk is blackbody}
\label{v592castable}
\end{deluxetable}

We decided to use this system to investigate the possibility of using the multi-wavelength astrometric capability of SIM Lite to discern two more subtle properties of disk-dominated systems: the disk temperature gradient and disk hotspots. The disk model employed in ELC allows for an inner disk temperature to be specified, along with a power law exponent, so that the disk temperature will vary radially according to the formula \begin{equation} T(r) = T_{inner}\cdot(\frac{r}{R_{inner}})^{\chi}, \end{equation} where T is the temperature at a given distance from the compact object, $r$, given a temperature T$_{inner}$ at the innermost radius, R$_{inner}$, and a power law exponent $\chi$ \citep{Orosz00}. Usually a value of $\chi$ = $-$0.75 is assumed for a steady-state disk \citep{Pringle81}, but we wanted to test if changing $\chi$ to $-$0.5 or $-$1.0, (which physically reflect a centrally irradiated disk \citep{Friedjung85,Vrtilek90,Bell99} and its corresponding extrema), would have a discernible effect on the wavelength dependent astrometric reflex motion. Thus, we produced two models, one with $\chi$ = $-$0.5 and the other with $\chi$ = $-$1.0, while also adjusting the inner disk temperature so that the resulting SED best matched that with $\chi$ = $-$0.75. Neither the SEDs nor the reflex motions differed significantly from the models that used $\chi$ = $-$0.75. Thus, it will not be possible to constrain the overall temperature gradient in accretion disks using astrometry.

\subsubsection{Sco X-1}

\begin{figure*}[t]
\centering
\begin{tabular}{cc}
\epsfig{file=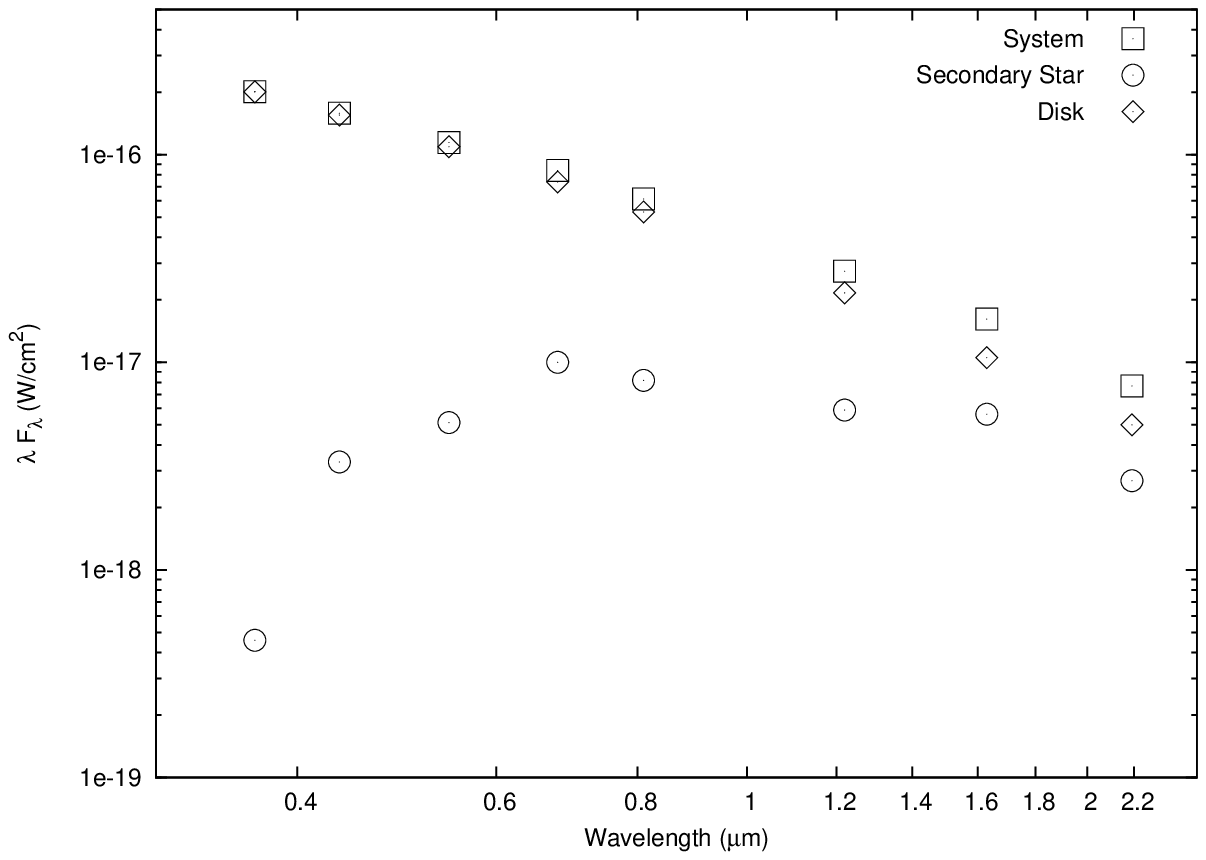, angle=0,width=0.45\linewidth} &
\includegraphics[width=0.33\linewidth]{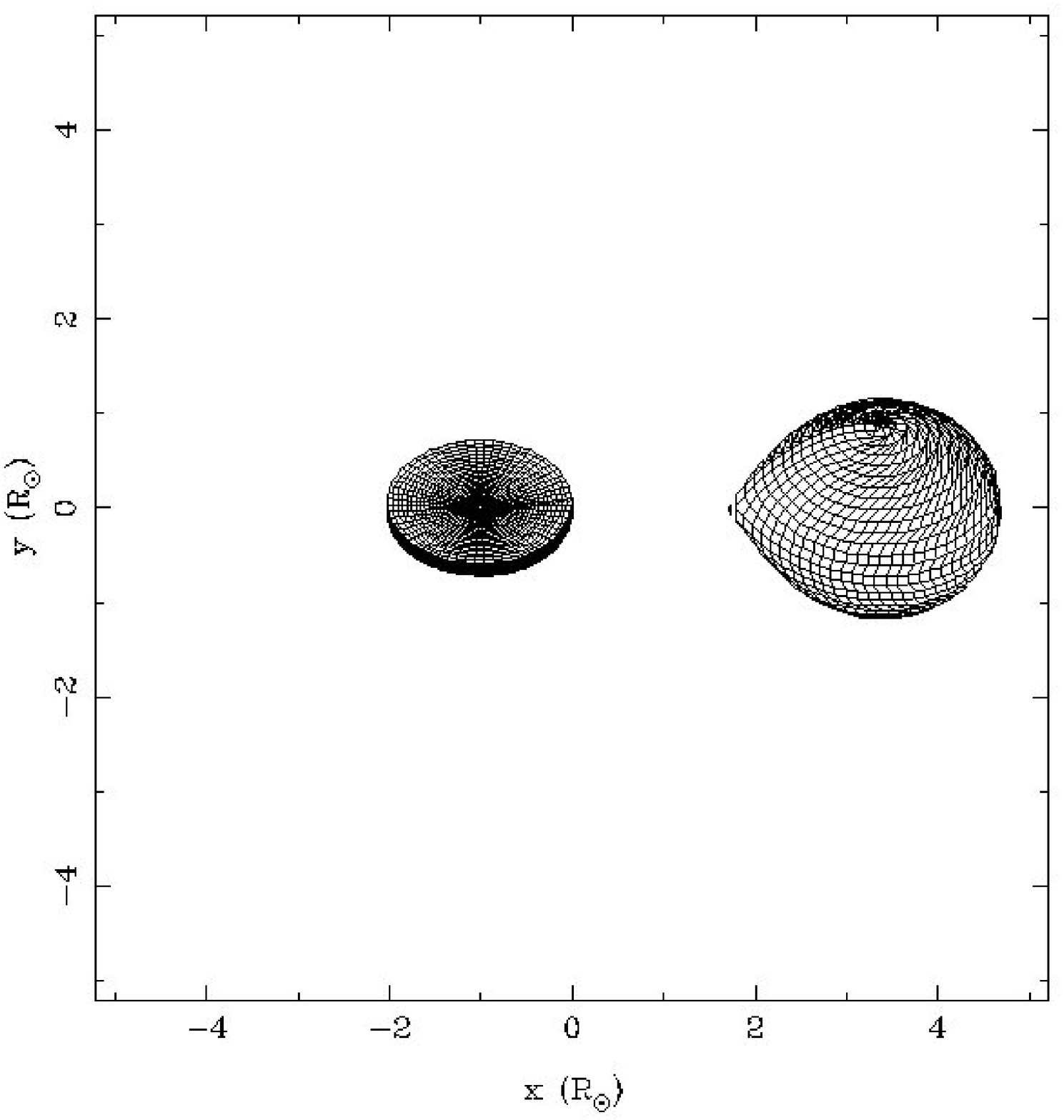}\\
\epsfig{file=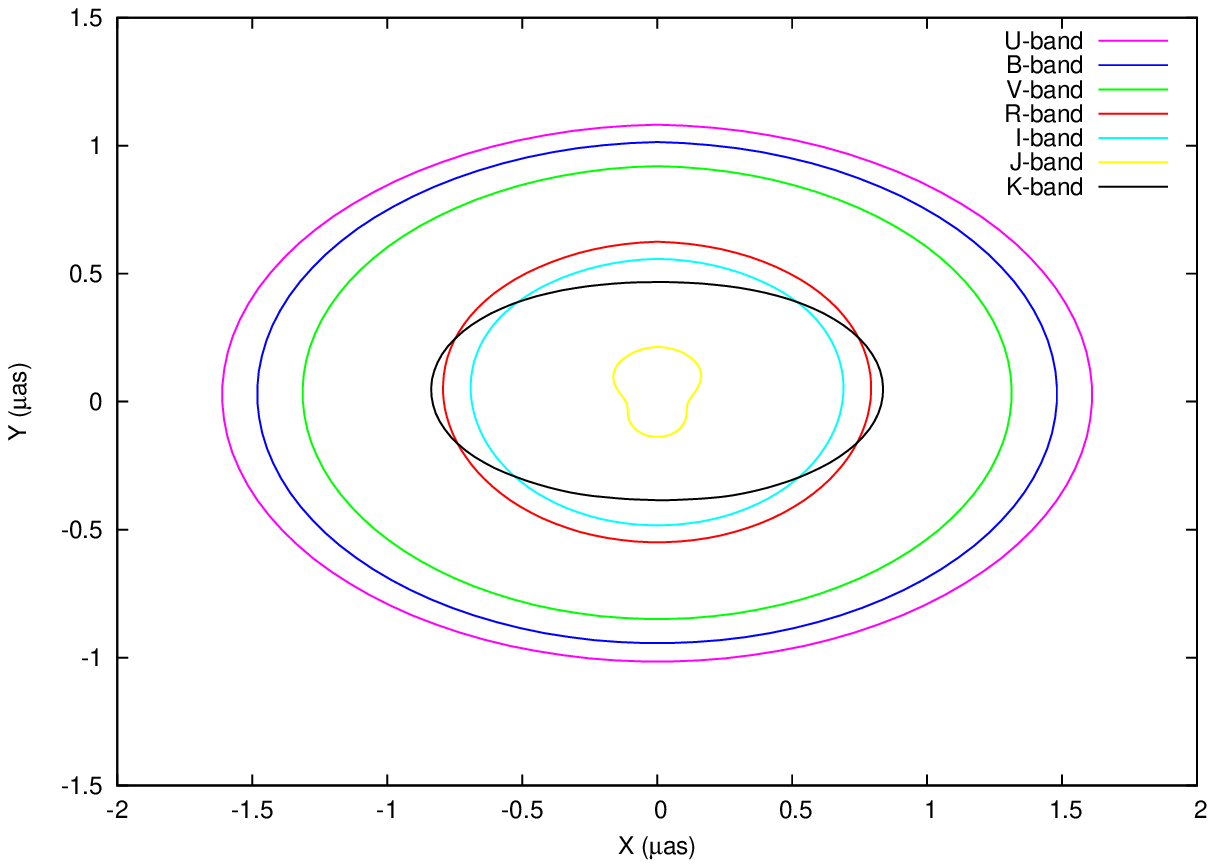, angle=0,width=0.45\linewidth} &
\epsfig{file=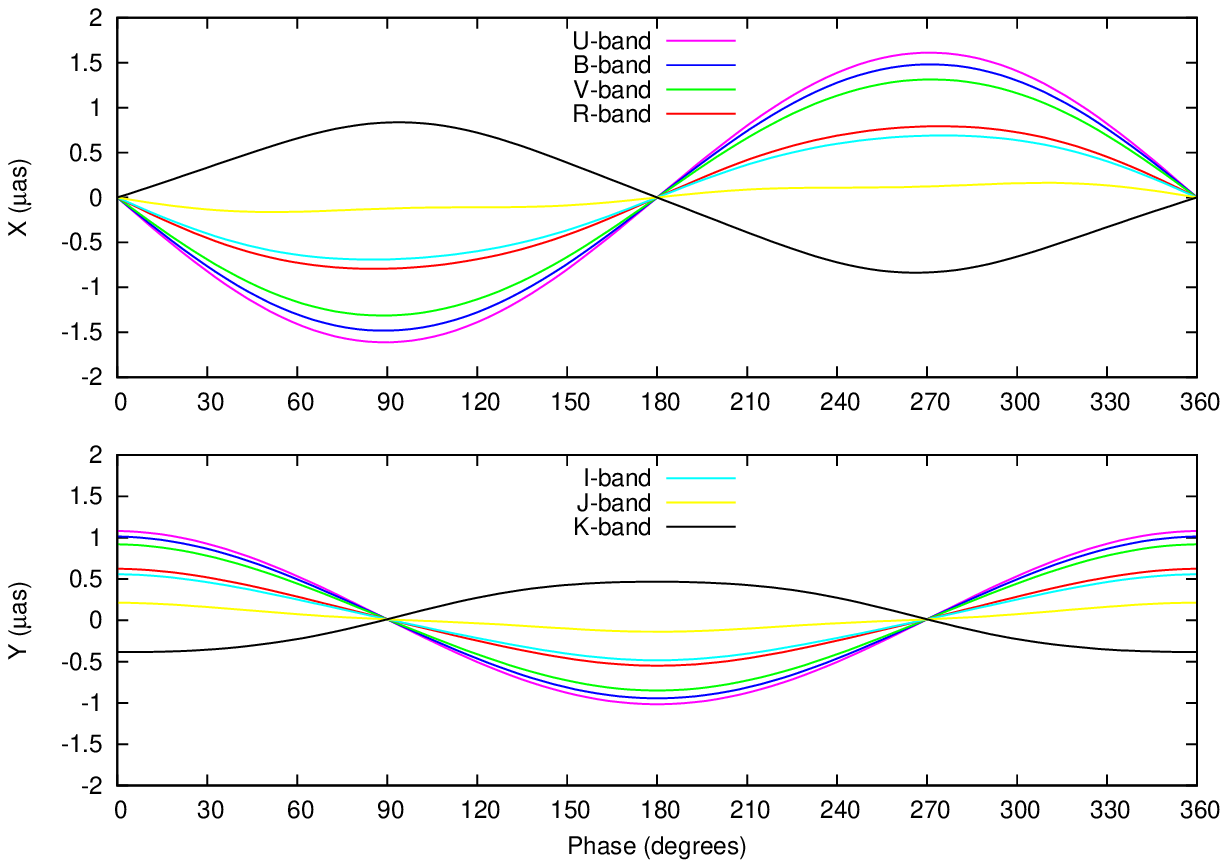, angle=0,width=0.45\linewidth} \\
\end{tabular}
\caption{\emph{Top Left:} SED plot for Sco X-1. \emph{Top Right:} 3D model of Sco X-1 at Phase=90$\degr$. \emph{Bottom Left:} Reflex orbit for Sco X-1. \emph{Bottom Right:} X and Y components of the reflex orbit versus phase for Sco X-1.}
\label{scox1mainfig}
\end{figure*}

\begin{deluxetable}{lc}
\tablewidth{0pt}
\tablecaption{Parameters for the Sco X-1 System}
\tablecolumns{2}
\tablehead{Parameter & Value\tablenotemark{a}}
\startdata
Magnitude (V) & 11.1 (max) 14.1 (min) \\
Distance (kpc) & 2.80\\
Inclination ($\degr$) & 50.0\\
Period (Days) & 0.7875\\
Eccentricity & 0.0\\
Mass of Star 1 (M$_{\sun}$) & 0.42\\
Mass of Star 2 (M$_{\sun}$) & 1.4\\
Radius of Star 1 (R$_{\sun}$) & 1.473\tablenotemark{b}\\
Radius of Star 2 (R$_{\sun}$) & 0.000\\
T$_{\rm eff}$ of Star 1 (K) & 4500\\
T$_{\rm eff}$ of Star 2 (K) & 0\\
Disk Inner Radius (R$_{\sun}$) & 0.026\\
Disk Outer Radius (R$_{\sun}$) & 1.011\\
Disk Inner Temperature (K) & 100 000\tablenotemark{c}\\
Disk Temp Power-Law Exponent & $-$0.75
\enddata
\tablenotetext{a}{Values from \citet{Steeghs02}, \citet{Gott75}, and \citet{las85}}
\tablenotetext{b}{Star fills its Roche Lobe}
\tablenotetext{c}{Disk is blackbody}
\label{scox1table}
\end{deluxetable}

In many CV systems, the optical light is orbitally modulated due to the presence of a hotspot on the outer edge of the accretion disk. This hotspot is where the accretion stream from the secondary impacts the disk. These spots typically have temperatures of 20,000 K, and can produce modulations on order of $\pm$50\% in visible bandpasses \citep{Mason02}. Because this feature is on the outer edge of the disk, it has the potential to distort the reflex motions in systems where it is prominent. In addition, due to its higher temperature than the surrounding disk, the hotspot could be the source of strong line emission, especially in HI or HeI lines \citep{Skidmore02}. Thus, its detection might be isolated, and the reflex motions amplified, using very narrow bandpasses centered on the strongest emission lines. To investigate the effects of a hotspot, we used the same V592 Cas model as above, but added a hotspot (20,000 K) on the outer edge of the accretion disk that leads the secondary by 30$\degr$ in phase, and that is 30\% of the $V$-band flux.  The differences between a model with a spot, and one without are extremely slight, with a maximum difference in models with and without a spot of 0.15 $\mu$as. Increasing the inclination of the system does not have a significant effect. To be even remotely detectable, V592 Cas would have to be $\sim$3 times closer, making the spot signature $\sim$0.5 $\mu$as, although it would still only be $\sim$5\% of the reflex amplitude signature. Unless an IB has a hotspot that is much more dominant, such features will remain undetectable.

\begin{figure*}[t]
\centering
\begin{tabular}{cc}
\epsfig{file=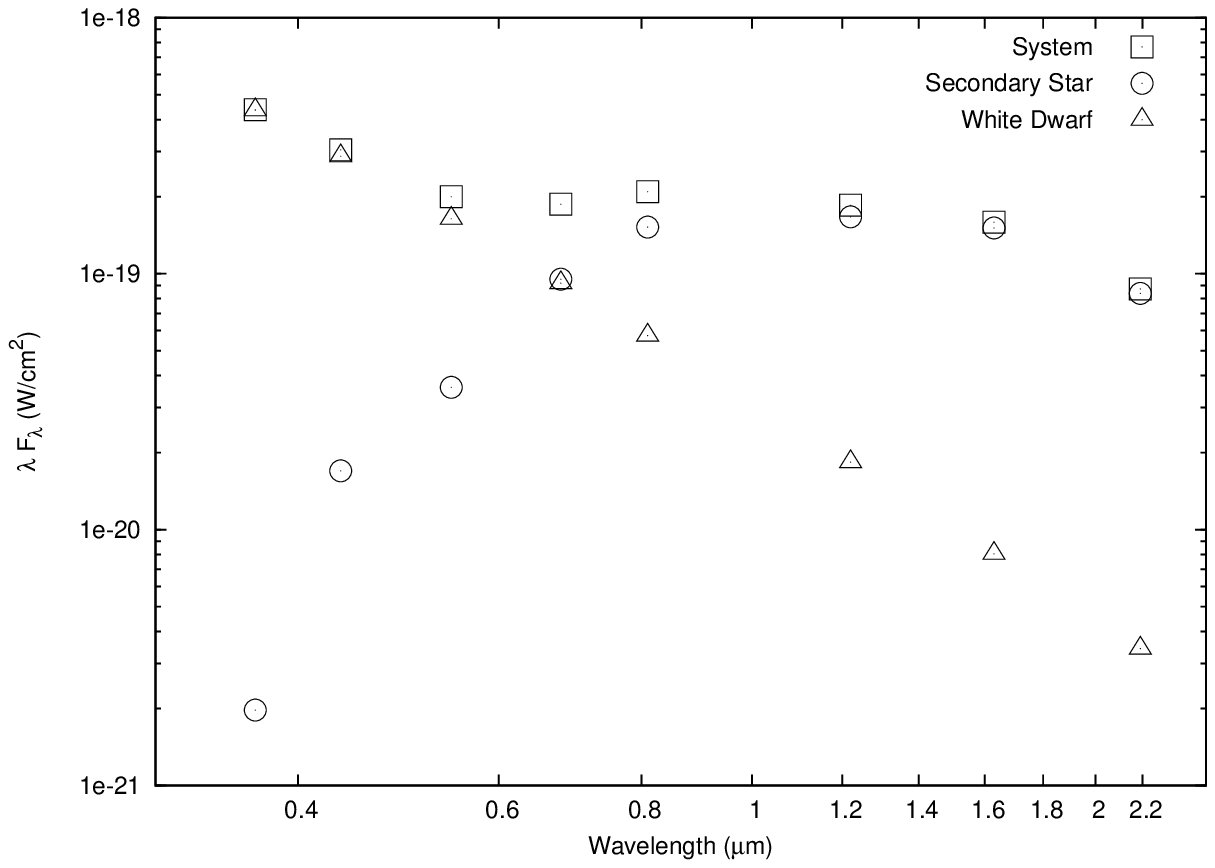, angle=0,width=0.45\linewidth} &
\includegraphics[width=0.33\linewidth]{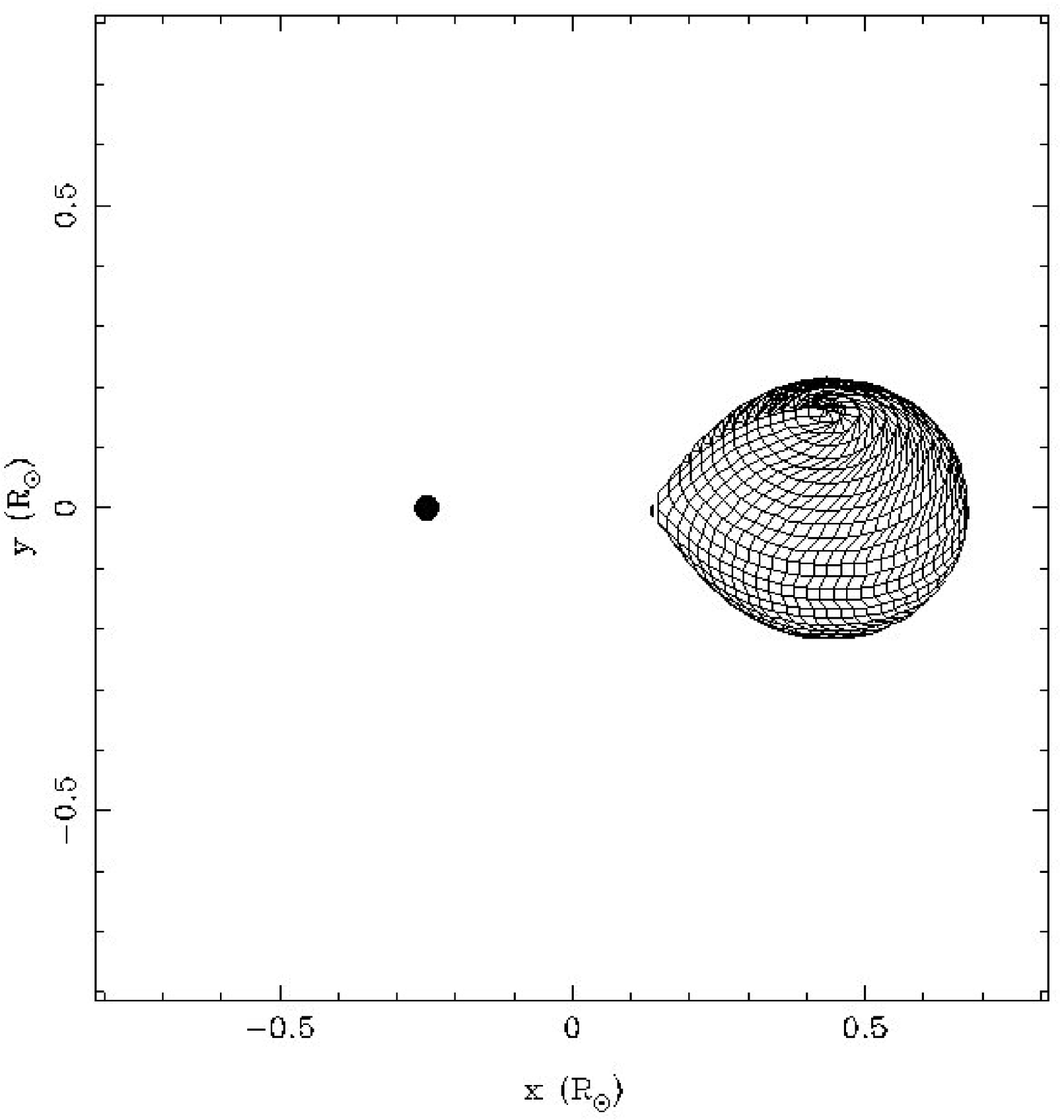} \\
\epsfig{file=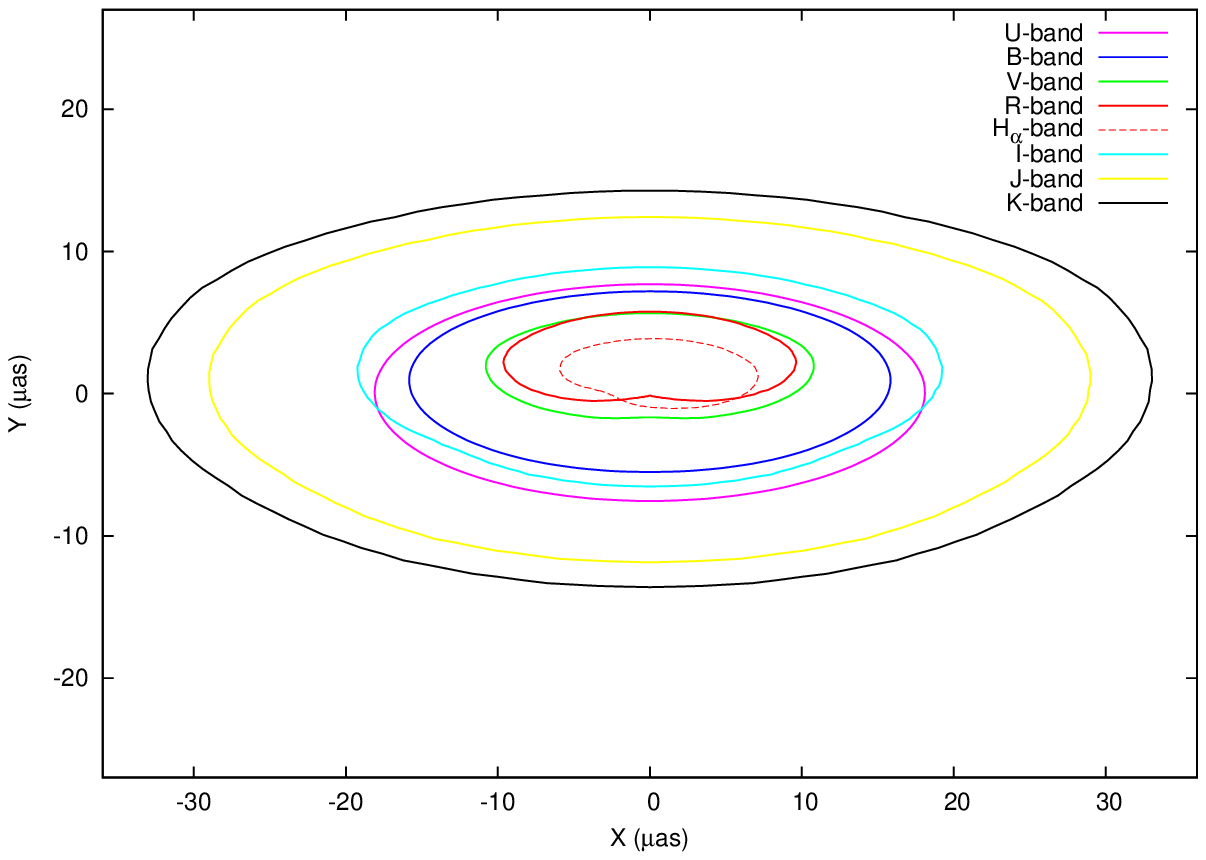, angle=0,width=0.45\linewidth} &
\epsfig{file=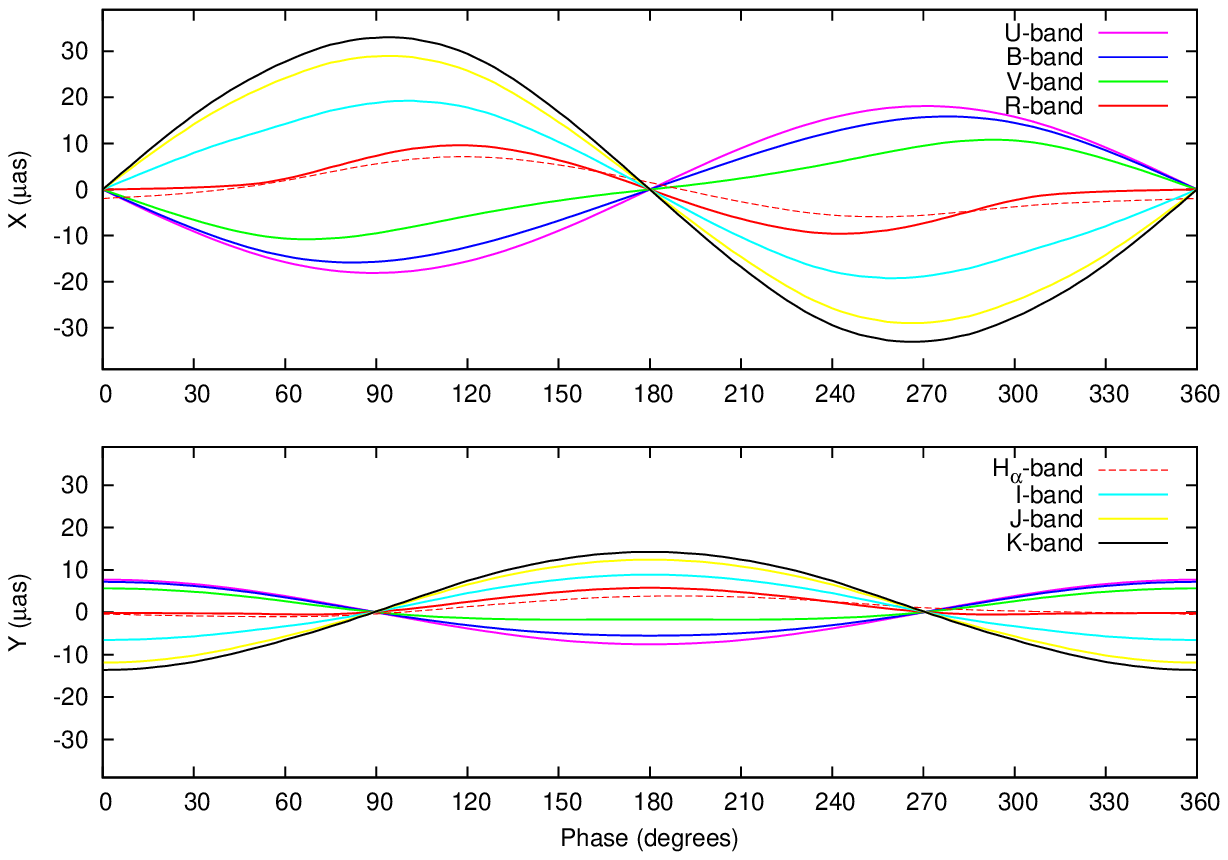, angle=0,width=0.45\linewidth} \\
\end{tabular}
\caption{\emph{Top Left:} SED plot for AR UMa. \emph{Top Right:} 3D model of AR UMa at Phase=90$\degr$. \emph{Bottom Left:} Reflex orbit for AR UMa. \emph{Bottom Right:} X and Y components of the reflex orbit versus phase for AR UMa.}
\label{arumamainfig}
\end{figure*}

Sco X-1 is the proto-type LMXB consisting of a neutron star accreting from a low-mass companion. The X-ray luminosity of Sco X-1 is very close to the Eddington limit for a 1.4 M$_{\sun}$ neutron star. The secondary star in Sco X-1 has never been directly detected, though an orbital period of 0.787 days was detected through the analysis of 85 years of visual photometry \citep{Gott75}, and from radial velocity measurements \citep{las85}. \citet{Steeghs02} detected narrow HI emission lines that they showed are from the irradiated secondary star. From these data they estimate masses for the components in this system of M$_{\rm 1}$ = 1.4 M$_{\sun}$ and M$_{\rm 2}$ = 0.42 M$_{\sun}$, thus making the secondary star a significantly evolved subgiant. Since the accretion disk dominates the spectrum, we have simply assumed a Roche lobe-filling M type secondary star corresponding to the observed mass, and an invisible neutron star. The stellar and accretion disk parameters are listed in Table~\ref{scox1table}, and the accretion disk model we employ is similar to that for V592 Cas. 

\begin{deluxetable}{lc}
\tablewidth{0pt}
\tablecaption{Parameters for the AR UMa System}
\tablecolumns{2}
\tablehead{Parameter & Value\tablenotemark{a}}
\startdata
Magnitude (V) & 14.50 (max) 18.00 (min)\\
Distance (pc) & 85.0\\
Inclination ($\degr$) & 65.0\\
Period (Days) & 0.0805\\
Eccentricity & 0.0\\
Mass of Star 1 (M$_{\sun}$) & 0.70\\
Mass of Star 2 (M$_{\sun}$) & 1.40\\
Radius of Star 1 (R$_{\sun}$)\tablenotemark{b} & 0.322\\
Radius of Star 2 (R$_{\sun}$) & 0.013\\
T$_{\rm eff}$ of Star 1 (K) & 3200\\
T$_{\rm eff}$ of Star 2 (K) & 35000
\enddata
\tablenotetext{a}{Values from \citet{Howell01a} and \citet{Harrison05}}
\tablenotetext{b}{Star fills its Roche Lobe}
\tablenotetext{c}{Disk is free-free}
\label{arumatable}
\end{deluxetable} 

As seen in Figure~\ref{scox1mainfig}, the accretion disk dominates until the near-infrared when the secondary star begins to contribute. Thus, as also seen in Figure~\ref{scox1mainfig}, the reflex motion has a weak wavelength dependence, but SIM Lite cannot observe in the near infrared where the reflex motion reverses, and thus it will be difficult to disentangle the components with such a weak wavelength dependence in the optical. However, it might be possible to recover the contribution to the astrometric reflex motion for systems with slightly more prominent secondary stars (e.g., Cyg X-2). As seen in Figure~\ref{scox1mainfig}, Cyg X-1 has a very small reflex motion, only $\sim$1.25 $\mu$as, and can get relatively faint, (V = 14.1 at its faintest), and thus this system is just at the limit of SIM Lite's capabilities. Adopting the faintest magnitude of 14.1 of the system to be conservative, and limiting ourselves to observations of 110 minutes, (0.1 of the orbital period), to prevent orbital smearing, we find via DAPE that we can achieve individual measurements with precisions of 2.2 $\mu$as, almost twice the semi-major axis of the system. Thus, assuming we would need at least 50 measurements to begin to derive a reasonable orbit for this system, this system would require at least 4 days of SIM Lite mission time, which may prove more expensive than the scientific payoff justifies. If we are able to observe in its high state when V = 11.1, we could achieve 30 measurements with 1.5 $\mu$as precision in only 20 hours of mission time, which is far more reasonable.  When varying the inclination of a Sco X-1 type system, the contribution from the secondary star becomes more significant as more of the disk becomes self-eclipsed and its total luminosity decreases, and thus alters the individual wavelength reflex motions accordingly.

\subsection{Polars and the Astrometric Signature of Magnetic Threading Regions}

\subsubsection{AR UMa}

AR UMa is a ``polar'', a CV with a highly magnetic WD primary and a Roche lobe filling M5/6 dwarf \citep{Harrison05}. As shown in \citet{Szkody99}, the $V$-band light curve is complex, being dominated by phase-dependent cyclotron emission. The near-IR light curve shows normal ellipsoidal variations, allowing \citet{Howell01a} to estimate an orbital inclination of 70$\degr$. The magnetic field strength in AR UMa has been estimated to be as high as B = 240 MG, though \citet{Howell01a} argue for a lower value of 190 MG. In these systems, at some point after material from the secondary star has passed through the inner Lagrangian point, it is captured by the magnetic field of the white dwarf, producing a ``magnetic threading region'' that can be a significant source of luminosity in H$_{\alpha}$. We investigate the possibility of astrometrically detecting this region via the following modeling process. We model the system as a white dwarf + red dwarf system without an accretion disk, using the parameters shown in Table~\ref{arumatable}, but for \emph{only} the  H$_{\alpha}$ bandpass add a narrow ``accretion'' disk halfway between the inner Lagrangian point and the WD. This disk then has a hot spot leading the inner Lagrangian point by 45$\degr$ in phase. The actual contribution from the disk has been made to be completely negligible, but the spot itself makes the system 25\% brighter in H$_{\alpha}$. In effect, it is an isolated emission spot located between the WD and secondary star, mimicking a threading region.

As can be seen in Figure~\ref{arumamainfig}, there is a large amount of  wavelength dependent astrometric reflex motion as expected, ranging from $\sim$10-20 $\mu$as in the optical, with the WD dominating at short wavelengths and the secondary star dominating at long wavelengths. This should easily allow the determination of both component masses with orbital coverage at a few passbands. According to DAPE, limiting individual observations to 12 minutes, (0.1 of the orbital period), to prevent orbital smearing, one could obtain individual measurements of $\sim$6 $\mu$as precision at the system's brightest, (V = 14.5), or $\sim$48 $\mu$as at its faintest, (V = 18.0). Thus, in its high state one could obtain 30, $\sim$6 $\mu$as precision, measurements in as little as 9 hours of total mission time, but in its low state, assuming 100, $\sim$48 $\mu$as precision, measurements would yield a meaningful solution, a total mission time of 30 hours would be required, which is still a reasonable amount of time for the scientific payoff in our opinion.

Of particular interest is that in H$_{\alpha}$ the threading region produces a detectable astrometric signature that has a phase offset of $\sim$15$\degr$ from the broadband wavelengths, due to the original 45$\degr$ offset of the hot spot diluted by the light from the rest of the system. Thus, through use of narrow-band astrometry, one could recover the location of the threading accretion regions in these types of systems. When varying the inclination angle for a AR UMa type system, the y component of the astrometric motion decreases with increasing inclination angle for all wavelengths as expected, but at large inclination angles eclipse effects produce significant deviations from simple sinusoidal reflex motions.

\section{Simulated SIM Observations}
\label{solvesec}

Although it is beyond the scope of this paper to perform a full simulation of exactly what parameters and to what precision one could extract with a given amount of SIM Lite time at specific wavelengths for each system, we can present a basic analysis for the SS Cyg system. To test the accuracy and precision to which we can recover system parameters, we generated simulated astrometric data for SS Cyg in the U and V bands, using the parameters listed in Table~\ref{sscygtable}, but with an inclination of 54.0$\degr$, comprised of 10 measurements at random phases with 1.7 $\mu$as Gaussian noise added to each measurement. We then simultaneously solved for the astrometric parameters of period, semi-major axis, inclination, the position angle of the line of nodes, and the angle in the plane of the true orbit between the line of nodes and the major axis, while fixing the the eccentricity and the epoch of passage through periastron to be zero, using the \textsc{GAUSSFIT} program \citep{Jefferys88}, with a procedure described in \citet{Benedict01}. We recover inclinations of 55.6 $\pm$ 5.1 and 54.0 $\pm$ 3.8 degrees, and semi-major axes of 13.58 $\pm$ 0.94 and 23.58 $\pm$ 1.28 $\mu$as, each for the U and V-bands respectively. Given the distance to the system of 159.5 parsecs \citep{Dubus04}, and assuming that 58\% of the U-band astrometric motion results from the white dwarf, and 69\% of the V-band flux results from the secondary star, as per the modeling, (see \S3.2.1), this yields masses of 0.559 $\pm$ 0.070 and 0.815 $\pm$ 0.096 M$_{\sun}$ for the secondary star and white dwarf respectively, compared to the input masses of 0.555 and 0.812 M$_{\sun}$. Thus, one will be able to disentangle individual component masses with reasonable uncertainties for this system with only $\sim$8 hours of SIM mission time.\\

\section{Conclusion}

We have presented a modeling code, \textsc{reflux}, that is capable of modeling the multi-wavelength astrometric signature of interacting binaries. We have presented models for multiple IB systems, and find that for primary or secondary star dominated systems, the contamination of the photocenter is minimal and SIM Lite will provide good astrometric orbits if the system is bright enough, yielding absolute inclinations and indirect masses. For mixed component systems, SIM Lite will be able to directly determine absolute masses for both components when multi-wavelength astrometric curves are obtained and combined with existing spectroscopy and multi-color photometry. For disk-dominated systems, SIM Lite will only be able to obtain astrometric orbits for a few systems, due to the small relative motion of the compact object that the disk surrounds, but these should yield accurate inclinations. We find that while multi-wavelength SIM Lite data will be unable to distinguish between various disk temperature gradients and disk hotspots, it should be able to determine the location of magnetic threading regions in polars, and thus similar effects in other systems, via narrow wavelength astrometry. In total, SIM Lite should contribute greatly to our understanding of interacting binary systems, especially if its multi-wavelength capability is maintained in the final design.

\acknowledgments
This work was sponsored in part by a SIM Science Study (PI: D. Gelino) from the National Aeronautics and Space Administration through a contract with the Jet Propulsion Laboratory, California Institute of Technology. J.L.C. acknowledges additional support from a New Mexico Space Grant Consortium Fellowship. The authors thank the referee for comments which helped to improve the manuscript.

\end{document}